\documentclass[conference]{IEEEtran}
\usepackage[nopdftrailerid=true]{pdfprivacy}
\usepackage[english]{babel}
\usepackage{amsmath}
\usepackage{amssymb}
\usepackage{booktabs}
\usepackage{comment}
\let\labelindent\relax \usepackage{enumitem}
\usepackage{graphicx}
\usepackage{subfig}
\usepackage{tabularx}
\usepackage{tikz}
\usepackage{xspace}
\usepackage{xurl}
\usepackage{array}
\usepackage{makecell}
\usepackage{multirow}
\usepackage{etoolbox}
\usepackage[htt]{hyphenat}
\usepackage{microtype}
\usepackage[subtle]{savetrees}
\setitemize{noitemsep,topsep=0pt,parsep=0pt,partopsep=0pt}
\usepackage[colorinlistoftodos]{todonotes}
\newcommand{\mypar}[1]{\medskip\noindent\textbf{#1}}
\newcommand{\datasetsCardinality}{$183,340$\xspace} % 100000+69700+3769+9871

\newcommand{\aptDatasetCardinality}{$3,769$\xspace}
\newcommand{\goodwareDatasetCardinality}{$9,871$\xspace}
\newcommand{\totTechniques}{$53$\xspace}
\newcommand{\ourTool}{\emph{Pepper}\xspace}
\begin{document}
\title{Longitudinal Study of the Prevalence of \\ Malware Evasive Techniques}
\author{%
  \IEEEauthorblockN{%
    Lorenzo Maffia\IEEEauthorrefmark{1},
    Dario Nisi\IEEEauthorrefmark{2},
    Platon Kotzias\IEEEauthorrefmark{3}
    Giovanni Lagorio\IEEEauthorrefmark{1}, 
    Simone Aonzo\IEEEauthorrefmark{2},
    Davide Balzarotti\IEEEauthorrefmark{2}
  }%
  \IEEEauthorblockA{\IEEEauthorrefmark{1} University of Genoa}%
  \IEEEauthorblockA{\IEEEauthorrefmark{2} EURECOM}%
  \IEEEauthorblockA{\IEEEauthorrefmark{3} NortonLifeLock}%
}
\maketitle
%! TEX root = ../main.tex
\begin{abstract}

By their very nature, malware samples employ a variety of techniques to
conceal their malicious behavior and hide it from analysis tools. To mitigate the problem, a
large number of different evasion techniques have been documented over
the years, and PoC implementations have been collected in public
frameworks, like the popular \emph{Al-Khaser}. As malware authors tend
to reuse existing approaches, it is common to observe the same evasive
techniques in malware samples of different families. 
However, no measurement study has been 
conducted to date to assess the adoption and prevalence of evasion techniques.

In this paper, we present a large-scale study, conducted by dynamically analyzing more than 180K Windows malware samples, on the evolution of evasive techniques over the years.
To perform the experiments, we developed a custom Pin-based Evasive Program Profiler (\ourTool),
a tool capable of both detecting and circumventing \totTechniques anti-dynamic-analysis techniques of different categories, ranging from anti-debug to virtual machine detection.

To observe the phenomenon of evasion from different points of view, we
employed four different datasets, including benign files, advanced persistent threat (APTs),
malware samples collected over a period of five years, and a recent
collection of different families submitted to VirusTotal over a one-month
period.
\end{abstract}
%! TEX root = ../main.tex
\section{Introduction} \label{sec:introduction}

The adversarial nature of malware analysis is ultimately the main aspect that distinguishes this field from other forms of program analysis. 
For instance, the widespread use of obfuscation and packing reduced the applicability of approaches based on static analysis and forced the security community to rely on a more costly form of dynamic analysis to observe the behavior of unknown samples.
As a natural reaction, malware authors immediately started to implement a multitude of tricks to detect different forms of dynamic analysis. 
This allows samples to hide their
malicious activity when they detect the presence of an analysis
environment, often resulting in an erroneous classification from the
malware analysis system. 

To mitigate this threat, researchers have followed three different approaches.
First, they have invested a considerable effort to \emph{document} malware evasion techniques as part of scientific papers, white-hat blog-posts, and open-source projects~\cite{branco2012scientific, barbosa2014prevalent, d2020dissection,lindorfer2011detecting, antiDebug, mitreProcessInjection, fireeyereverseturin, mcafee2020evolution}.
On the one hand, practitioners can benefit from this information to identify and deal with evasive malware.
On the other hand, malware authors also take advantage of publicly documented evasion techniques, which allow even non-experts to incorporate sophisticated evasion techniques in their samples.
In fact, as widely reported in the literature, malware authors are no stranger to code-reuse and cut\&paste approaches~\cite{hanna2012juxtapp, codereuse, cozziiotmalware}.

A second line of research has focused on \emph{detecting} evasive
samples~\cite{d2019sok,polino2017measuring}, to trigger more in-depth
analysis phases or to repeat the execution in a different environment
(e.g., in a bare-metal computer instead of a virtual machine).

Finally, a third line of research in this field has focused on \emph{mitigation}.
In this case, researchers have tried to devise new dynamic analysis
techniques that are more difficult to detect, by either moving the
instrumentation out of reach from the running
sample~\cite{dinaburg2008ether,fattori2010dynamic,kirat2011barebox} or by
hiding the instrumentation by providing fake runtime information to the
program~\cite{polino2017measuring,d2019sok,d2020dissection}.

Overall, the three aforementioned aspects of discovery, detection, and mitigation have been widely covered in the research literature. 
However, to date, no comprehensive study has been conducted to
\emph{measure} the adoption and prevalence of different evasion techniques in the
malware ecosystem.
In other words, while it is clear that malware authors routinely employ tricks to evade dynamic
analysis, it is still unclear which techniques are used more frequently, how many tricks are implemented by single sample, 
in which order, and at which point in time during the program execution.
We believe that a precise measurement of these factors is paramount to support further research in the field and to properly choose and configure existing solutions.
In fact, mitigations for many classes of anti-analysis techniques are time-consuming to develop and costly (in terms of overhead, thus limiting scalability) to deploy.

The second goal of our study is to measure the evolution of evasive samples
over time and to compare this phenomenon across different representative datasets that contain different types of samples.
Finally, we plan to investigate the relationship between the presence of anti-analysis techniques and the use of packers or protectors.

The main challenge we encountered to answer these questions is that
conducting the experiments require a measurement tool capable not only of
detecting but also of \textit{mitigating} all known evasion techniques against dynamic analysis.
This is required in order to deceive malware samples in continuing
their execution and potentially exhibit more anti-analysis techniques.
For this reason, we developed \ourTool a Pin-based Evasive Program ProfilER based on the Intel Pin~\cite{intelpin} dynamic binary instrumentation (DBI) framework. 
\ourTool can precisely identify 53 different techniques,
divided in seven categories -- which corresponds to \emph{all} the evasive solutions 
publicly documented to date.

By using our tool, we measured the prevalence of evasion techniques on four
different datasets of Windows executables, constructed with different
criteria to either cover multiple families or multiple years of appearance,
as well as including samples from advanced persistent threats and benign
applications. In total, we dynamically analyzed \datasetsCardinality
samples, the largest dataset to date used in this area. This large-scale
investigation allowed us to measure
the evolution of the prevalence of evasion techniques throughout the years,
characterize their adoption among different classes of malware, and compare
their behavior against a control group.

We found that malware has increased its evasive behavior over the years, with more than 40\% of the samples collected in 2020 employing at least one technique, in comparison with about 30\% in 2016.
We also found a certain degree of diversity in the number of techniques each sample employ:
while, on average, each sample uses two techniques, some go as far as implementing more than ten.

The use of benign samples allowed us to identify four techniques that,
while often reported as anti-analysis tricks, are also commonly used for
legitimate purposes.
We discuss these possible `false positives' in details, and use our results
to show how the presence of these techniques can not be reliably used as an
indicator of evasive behavior.

We also noticed how the evolution of anti-analysis techniques has changed over the years.
On the one hand, VM Checks (in short, when malware checks for the presence
of an artifact ascribable to a virtualized environment) have decreased,
probably because server virtualization has become the standard in the
enterprise world; thus, not infecting virtualized machines would be
counter-productive for malicious actors. 
On the other hand, the use of Anti-Debugging
techniques has increased, presumably to account for the increasing number of cybersecurity
professionals with reverse-engineering skills.
To put this into numbers, about 20\% of recent malware samples use a VM-Detection technique, while 80\% employ Anti-Debugging tricks.

Moreover, our analysis allowed us to precisely measure \emph{when} evasive
behavior occurs during malware execution: as one may expect, the majority
of samples exhibit it at the beginning of their execution. However, we also
found a non-negligible amount of samples that use evasive techniques
throughout their entire execution. The order is also important, as anti-debugging is
almost always used \textit{before} other anti-analysis techniques.

Finally, we studied the impact of packers and
protectors on dynamic analysis and found that samples that use protectors are more likely to combine
multiple evasive behaviors. In contrast, packers focus mainly on
removing the PE header from memory to prevent the analyst from dumping
their unpacked code.

In the spirit of open science, and to allow our results reproducibility, we release\footnote{\url{https://github.com/Maff1t/JuanLesPIN-Public}} the source code of \ourTool and the hashes of the samples we analyzed in our experiments.

%! TEX root = ../main.tex
\section{Background and Related Work}
\label{sec:background}

Dynamic malware analysis consists of running a suspicious application inside a controlled and instrumented environment to observe and collect different aspects of the program behavior. 
The data collected by the system can then be used to look for evidence of malicious behavior, extract artifacts and indicators of compromise, or classify the sample in a particular class or malware family.

A large body of research exists on the design and implementation of dynamic malware analysis systems~\cite{bayer2006dynamic,willems2007toward,yin2007panorama,dinaburg2008ether,pek2011nether,kirat2011barebox,lengyel2014drakvuf,kruegel2014how,henderson2016decaf,polino2017measuring,cozzi2018understanding}.
In particular, researchers have focused on two main aspects of dynamic analysis: the \emph{runtime environment} and the \emph{analysis component}.
The first includes the entire hardware and software stacks that make the execution of malware samples possible in the first place.
In order to isolate the execution of untrusted code and improve the infrastructure scalability, 
the majority of malware analysis systems are implemented through Virtual Machines (VM) by using either a virtualized~\cite{pek2011nether, lengyel2014drakvuf} or an emulated~\cite{bayer2006dynamic,yin2007panorama, henderson2016decaf, cozzi2018understanding} hardware stack.
However, the advantage in terms of scalability comes at the cost of stealthiness as several aspects of the underlying virtual environments are fundamentally different from those found in the average end-user machine, thus making VM detection a well-known target for malware developers to evade the analysis environment.
Bare-metal runtime environments~\cite{kirat2014barecloud} try to circumvent such evasion techniques by employing the same physical machines an end-user would use -- at the cost of reduced scalability and more complex instrumentation.

The analysis component covers instead the system instrumentation responsible to capture the behavior of the investigated sample.
In this case, different solutions offer a trade-off between stealthiness and both the amount and the granularity of the information collected by the system.
A coarse-grained classification divides instrumentation placed within the same operating system instance of the analyzed sample (an \emph{in-guest} component) from those placed in a separate instance (\emph{out-of-guest}).
In-guest components can run both in userspace, as it is the case for debuggers and Dynamic Binary Instrumentation (DBI) tools (e.g., Intel Pin~\cite{intelpin}, and DynamoRIO~\cite{bruening2012transparent}); or as part of the OS kernel, like in the case of kernel modules that provide syscall monitoring capabilties~\cite{bayer2006dynamic}.
Out-of-guest components are instead usually implemented as part of the virtualized or emulated runtime environment (e.g., PANDA~\cite{dolan2015repeatable}, and PyREBox~\cite{pyrebox}) or as a separate entity like a network packet inspector.

In-guest components are usually more invasive but provide more semantic-rich information about the malware behavior.
DBI tools, for example, inject custom code within the context of the application, to monitor its execution flow 
% The application's overall behavior is guaranteed to be preserved, and by running within the context of the application, 
and access fine-grained information about the program's internal state, such as the list of invoked library calls and the executed basic blocks.
On the other hand, out-of-guest analysis components are generally less invasive but more difficult to detect.
As previous work has shown~\cite{nisi2019exploring}, the semantic gap between the information provided by in-guest and out-of-guest is hard to bridge, further limiting the adoption of the latter in malware analysis pipelines.

The variety of techniques discussed so far lead malware authors to introduce a large arsenal of techniques to evade dynamic analysis. 

\subsection{A Taxonomy of Anti-Dynamic Analysis Techniques} \label{subsec:taxonomy}

The first attempt to categorize evasion techniques used by malware dates back to 2008, when Chen et al.~\cite{chen2008towards} proposed the first taxonomy of known techniques.
More recently, Afianian et al.~\cite{afianian2019malware} revised the list and proposed a new classification of the evasion tactics and techniques found in modern malware.
For our study, we explored a large number of publicly documented dynamic-analysis evasion techniques, as described in white and scientific papers~\cite{branco2012scientific,barbosa2014prevalent,d2020dissection,lindorfer2011detecting}, as well as in online blogs and resources for ethical hacking~\cite{processInjection,antiDebug,mitreProcessInjection,fireeyereverseturin,mcafee2020evolution}.
A reference implementation of all known techniques is also provided by the recent and still actively maintained Al-Khaser~\cite{alkhaser} framework. 
Following Al-Khaser classification, we can distinguish seven evasion technique categories:

\mypar{Anti Debug.} Debuggers are commonly used for manual and automatic dynamic malware analysis, as they allow for a total control over the analyzed program.
Anti-debugging techniques consist of finding the many artifacts that debuggers introduce 
within the context of the analyzed program, including CPU registers and in-memory data structures.
The Windows operating system also provides specific APIs, such as \texttt{IsDebuggerPresent}, that malware samples can leverage for anti-debug purposes.
Another example is provided by the \texttt{NtSetInformationThread} native API, which supports an undocumented flag
\texttt{THREAD\_INFORMATION\_CLASS::ThreadHideFromDebugger} to hide a specific thread from a debugger.

\mypar{Anti Dump.} Malware analysts and incident-response practitioners often resort to memory forensic techniques
to find evidence of malware infections and to analyze 
file-less malware or samples that adopt custom packing schemes.

Anti-dump techniques hinder memory forensics by tampering with the data structures that standard tools used for their analysis.
For example, by erasing the PE header from the process virtual memory or by
increasing the value of the \texttt{SizeOfImage} field (in the
\texttt{IMAGE\_OPTIONAL\_HEADER}), a sample can
prevent Volatility~\cite{volatility} from correctly identifying the process
in memory, thus evading any further memory investigation. 

\mypar{Anti Instrumentation.} Covered in detail by Polino et al.~\cite{polino2017measuring}, these techniques aim to detect the presence of a DBI tool at runtime.
DBI tools employ Just In Time (JIT) compilers to instrument the original code of the analyzed binary at runtime.
Since the since the size of the instrumented code differs from the that of the original application, DBI tools cannot modify it in place
but need to rely on code caches, i.e., areas of memory in which the instrumented code is stored and from which it is executed.
This results, for instance, in a discrepancy in the value of the program
counter register between a normal and an instrumented execution of the same
program and malware can exploit this discrepancy for evasion.

\mypar{Code Injections.} Malware samples, especially file-less ones, often run their malicious code from within the context of other processes.
This strategy aims at evading defense mechanisms (e.g., antiviruses) and dynamic malware analysis tools, which often apply their analysis only to untrusted processes. 
By leveraging Windows API, malware can either directly inject its shellcode into another process or force it to load a malicious DLL.
Code injection techniques are also used for persistence.
For instance, malware samples can tamper with the \texttt{Appinit\_Dlls} registry key to force other processes to load malicious libraries, even after a reboot.

\mypar{Resource Profiling.} 
Malware can profile the available resources (e.g., the RAM size, the amount of disk space left, the number of processors, and the screen size) of the machine onto which it executes and infer, from this information, to which computer class this machine belongs.
For example, a large amount of RAM and disk space are typically found in server machines, while the presence of a large screen likely indicates a desktop computer.
On the contrary, to cope with the ever-increasing number of malicious samples, dynamic-analysis pipelines need scalability, which they achieve by limiting the hardware resources allocated for each sandbox instance.
For example, some sandbox solutions use a single processor, which is very uncommon in modern computers.
The downside is that many resource profiling techniques are also commonly used among goodware.
For instance, the number of processors is often used to allocate a process/thread pool, and the screen size can help a program to decide how to layout elements in GUI applications.

This category also includes techniques that scan the list of running processes to 
discover the presence of analysis tools, e.g., debuggers, packet sniffers, antiviruses, or guest tools in support of the hypervisor.
A sample can also exactly tell that a debugger or a DBI is analyzing it by checking its parent process.

\mypar{VM Checks.} Virtualization and emulation solutions aim at mimicking the functionalities of real hardware.
Despite being often advertised as entirely transparent from the guest programs' point of view, both hypervisors and emulators introduce artifacts that betray their presence.
Researchers have documented a long list of such artifacts, including specific files, processes, registry keys, services, and network adapters.
Malware authors try to retrieve this information in the most imaginative ways, 
such as by querying the Windows Management Instrumentation (WMI) to obtain the network adapters' MAC address.
Another example is the use of the \texttt{cpuid} instruction, which loads
various details about the machine's processor, including the CPU
manufacturer-ID string that, under some hardware virtualization solution,
reveals the hypervisor name and vendor.

\mypar{Timing Attacks.} Time-based techniques fall into two categories: time stalling and runtime measurements.
Stalling techniques exploit the fact that sandboxes are programmed to analyze samples for a certain amount of time.
Malware that employs stalling techniques pauses their execution for a certain amount of time to outwait the sandbox. 

Runtime measurement techniques exploit instead the fact that analysis components and runtime environments introduce
a noticeable overhead to specific operations (in some corner cases, they can
be even faster).
Malware authors can measure how much time a reference bare metal machine needs for executing a particular operation (e.g., an API call or a group of instructions) and check if the same operation takes a similar time on the machine onto which their malware runs.
The most common timing check method uses the \texttt{rdtsc} instruction~\cite{oyama2019does}, which returns the number of ticks since the last CPU reset.
Malware executes this instruction twice, before and after a certain amount of code, and compares the difference between the two readings with the same snippet executed on bare-metal.
For example, some instructions like the \texttt{cpuid} instruction, explained earlier, are slower on a hypervisor because of the trap-and-emulate virtualization: the execution of a privileged instruction by the guest operating system causes a trap, allowing the hypervisor to emulate the effects of that specific instruction.

It is worth noting that such runtime measurements are also closely related to the categories of VM Checks, Anti Instrumentation, and Anti Debug because all the underlying technologies introduce a non-negligible overhead.
However, we used a dedicated category because, in most cases, it is difficult to tell which technology the malware is trying to evade.

\subsection{Off-The-Shelf Packers and Protectors}\label{subsec:packers_protectors}

Both runtime packers and software protectors modify a program 
to make it more difficult to analyze while still preserving its original behavior.
Packing consists of compressing the code of an executable and later de-compressing it at runtime before its execution.
While originally packers were introduced to save disk space, they quickly became very popular among
malware authors for their ability to prevent static analysis.

As the name suggests, software protectors are more general tools that aim at protecting an executable from tampering and reverse engineering.
To achieve this, protectors often combine several techniques, including packing, encryption, and code virtualization.
The latter -- which is not to be confused with hardware virtualization -- consists of rewriting the code of an application by translating it into a virtual instruction set architecture (ISA).
The software protector embeds a software component, the so-called ``virtual machine'', that interprets the translated code at runtime.
The gaming industry adopts software protectors for anti-piracy and intellectual property protection purposes.
Their characteristics, however, made them viable tools for malware authors too, since they are explicitly designed to thwart reverse engineering.
Software protectors are often customizable, and many of them also provide a wide range
of evasion techniques against dynamic analysis.

While packers and protectors are often considered together, 
in our work, we decided to keep the two categories separate because of their orthogonal approaches.
In fact, while packers only transform applications without introducing additional checks (thus, for our purposes, crypters, as defined in~\cite{malwarebytespacker}, work similarly to packers), one of the protectors' main goals is to protect an application against dynamic analysis by applying evasive techniques.

\subsection{Analysis of Evasive Malware}

Evasive malware often adopts a strategy commonly referred to as ``split personality,'' which consists of exhibiting different behaviors in different environments~\cite{balzarotti2010efficient}.
The most notable application of this strategy is to thwart dynamic analysis by showing a benign behavior (or no behavior at all) when the malware finds evidence of an instrumented environment.
A common approach to detect split personality in malware consists of executing it in different analysis environments and comparing the observed behaviors~\cite{chen2008towards,balzarotti2010efficient,kirat2014barecloud}.
Previous work has adopted this approach also for discovering previously unknown evasion techniques~\cite{lindorfer2011detecting}, automatically creating behavioral signatures for evasion techniques~\cite{kirat2015malgene}, and mitigating anti-emulation techniques~\cite{kang2009emulating}.

Another line of research has focused on proposing techniques to mitigate or counter evasive malware.
These broadly fall into two categories: 1) those that aim at reducing the number of artifacts that the malware can use to detect the analysis environment, and 2) those that actively neutralize known evasion techniques at runtime.
Ether~\cite{dinaburg2008ether} was a seminal work in which the authors tried to achieve stealthiness by moving the analysis system in the hypervisor.
Fattori et al.~\cite{fattori2010dynamic} followed the same rationale in HyperDbg, a debugger implemented by using only virtualization primitives.
However, virtualization was not a perfect solution as Hypervisors also introduce artifacts that malware can leverage for detection.
The elimination of any artifact that virtualization introduces was what urged Kirat et al.~\cite{kirat2011barebox} to use unvirtualized (bare metal) machines in their design of BareBox.

In the second category, Polino et al. proposed \emph{Arancino}~\cite{polino2017measuring}, a system to hide the artifacts introduced by DBI, which effectively allows this technique to be used for malware analysis. The authors also measured how malware detects DBI tools by analyzing a dataset of 7,006 samples.
On the other hand, \emph{BluePill}~\cite{d2020dissection} handles evasive malware samples by forging the information used by several classes of operations commonly used for evasion.
For such a purpose, BluePill uses the mitigations techniques described in~\cite{d2019sok}.

For our study, we developed a tool inspired by Arancino and BluePill.
However, while these two previous works focused exclusively on the design of the anti-evasion techniques, our primary goal is to perform a large-scale measurement to assess the prevalence of evasive malware in the wild. 
For this reason, our tool requires a more fine-grained detection mechanism to reduce false positives and to pinpoint which evasion techniques the malware under analysis employs.
Moreover, it covers a broader range of techniques, such as process and DLL injection, that were not considered in previous studies.

\subsection{Prevalence of Evasive Malware}
In the past, both academia and industry have tried to estimate the spread of few selected dynamic analysis evasion techniques among malware samples, often reaching contradicting results.

In 2008, Chen et al.~\cite{chen2008towards} estimated the prevalence of
both anti-virtualization and anti-debugging techniques by running their
samples with and without instrumentation.
They measured that 40\% of the samples in a dataset of 6900 malicious programs exhibited less malicious behaviors when instrumented.
They also underlined that malware was more likely to conceal its behavior while being debugged (39\% of samples) than when running in a virtual machine (4\% of samples). 
The following year Bayer et al.~\cite{bayer2009view} found that only 0.03\% of the 901,294 samples submitted to the Anubis sandbox contained evasion techniques specifically tailored against their sandbox. The authors also reported that 12.45\% of their samples terminated without showing sufficient activity, a possible sign of evasive behavior. 

Since then, evasion techniques seem to have gained traction in the malware landscape.
In particular, Lindorfer et al.~\cite{lindorfer2011detecting} reported that 25.56\% of the samples in their dataset attempted to evade the Anubis sandbox.
In 2012, Branco et al.~\cite{branco2012scientific} studied the evasive behaviors of 883 samples and concluded that anti-debug and anti virtualization techniques were employed by 43.21\% and 81.40\% of the samples, respectively.
Lastly, in a report from Symantec~\cite{symantecvmdetect}, the authors reported that 20\% of samples collected from January 2012 to February 2014 adopted anti virtualization techniques targeting the VMWare hypervisor only.
The most recent study on the topic was conducted by Polino et al. ~\cite{polino2017measuring} in 2017. The authors focused on anti-instrumentation techniques, particularly those targeting DBI, and measured that 15.6\% of the 7,006 samples they analyzed implement at least one evasive technique.

Note how all these previous attempts used very small dataset and often unclear methodologies based on heuristics on the observed behavior.
As a result, experiments (even conducted on the same year) reached very different conclusions about the prevalence of anti-dynamic analysis techniques.

Our work offers more precise measurements and new insights, by performing
the first longitudinal study on samples collected over a period of five
years. Our dataset is between two and three orders of magnitude larger and our system
is able to capture 53 techniques divided into 7 categories.
In comparison, Polino~\cite{polino2017measuring} studied 5 techniques in a single category, and Branco~\cite{branco2012scientific} 3 categories.

%! TEX root = ../main.tex
\section{Measurement System}
\label{sec:tool}

To perform our experiments, we implemented a custom dynamic analysis tool named \ourTool: Pin-based Evasive Program ProfilER.
The system was designed with three main objectives.
First, it should be capable of detecting the presence of \emph{all} known evasive techniques used by malware authors against dynamic analysis.
Since analyzing evasive malware is, by definition, a challenging task and malware samples are usually obfuscated, often through custom packing schemes or virtual-machine-based obfuscators, static binary analysis is not a viable option.
For this reason, we decided to implement our detection routines by dynamically instrument the target binaries at runtime.

The second objective of our system is to counter each evasion technique by employing specific mitigation solutions.
In fact, in order to measure the prevalence of each evasion technique, detecting their presence is not sufficient.
As we mentioned in Section~\ref{sec:background}, malware often employs entire suites of evasion techniques and, as soon as one triggers, a sample could stop its execution or avoid triggering its malicious behavior.
Therefore, by relying on detection mechanisms alone, it is only possible to find evidence of at most one evasion technique per sample, 
thus making the investigation shallow and incomplete.
By adopting counter-evasion techniques, we can instead observe the complete list of evasive techniques employed by each sample.
Since we want to analyze hundreds of thousands of samples, the third and last objective of our system is to be scalable.

Finally, it is crucial to understand that our goal is to study the prevalence of each \emph{known} technique and not of evasion in general (which could be better detected, as previous works have shown, by running the sample in multiple environments).
Thus, our goal is not to design a new dynamic analysis system that is impossible to detect, nor to identify previously-unknown techniques.
We are also aware of the fact that malware authors, by studying our implementation, could find ways to evade our systems. 
However, this has no impact on our experiments and on the results presented in this paper.

\subsection{General Design}

We decided to implement \ourTool on top of Intel Pin~\cite{intelpin}, a popular DBI framework.
Using the Intel Pin APIs, one can instrument several aspects of a program's execution by re-routing them through custom procedures.
\ourTool is implemented in about 5000 lines of C++, and it leverages the flexibility it inherits from Intel Pin for the following four purposes.

\mypar{Evasion Detection.} \ourTool provides a set of routines that are executed when the program invokes certain Windows APIs, executes specific CPU instructions, or accesses certain areas of memory.
These are all events that, in one way or another, evasion techniques can exploit to detect the presence of an analysis environment.
For example, \emph{IsDebuggerPresent}, \emph{Interrupt\_3}, and \emph{IsDebuggerPresentPEB} are three techniques which leverage different types of artifacts to understand whether the current process is being debugged.
Our system detects each of them by hooking the corresponding event: for \emph{IsDebuggerPresent} the event is the execution of the homonym API; for \emph{Interrupt\_3} it is the execution of the \texttt{int 3} processor instruction, and for \emph{IsDebuggerPresentPEB} the triggering event is a memory read of a specific field within the Process Environment Block (PEB).

\mypar{Evasion Mitigation.} \ourTool hinders the evasion techniques that the malware put in place by actively changing the information that the malware extracts from the evasion artifacts.
For instance, the \emph{cpuid\_is\_hypervisor} leverages the instruction \texttt{cpuid} which, among other information, reveals whether a hypervisor is running on the machine.
\ourTool mitigates this technique by switching the corresponding bit in the right CPU register after a \texttt{cpuid} instruction is executed.

\mypar{Behavior Analysis.} \ourTool also collects information about the behavior of the analyzed sample by instrumenting sensitive Windows APIs.
In particular, \ourTool traces the same APIs that \emph{Cuckoo Sandbox}~\cite{oktavianto2013cuckoo} does.
These include network- and filesystem-related Windows native APIs, as well as those that manipulate processes, threads, and registry keys.
This allows our system to replace a traditional malware analysis sandbox and collect information about the runtime behavior of a sample.

\mypar{Logging Infrastructure.} We provided \ourTool with extensive logging capabilities through which it reports every monitored event.
\ourTool generates logs for the initial process in which the sample runs, its entire hierarchy of child processes, 
and for each external process in which code is injected at runtime.

%! TEX root = ../main.tex
\begin{table}[t]
\centering
\caption{Techniques per Category}
\label{tab:summarytechniques}
\small
\begin{tabularx}{\linewidth}{c l c c }
\toprule
\textbf{Table} & \textbf{Category} & \textbf{\# of Techniques} & \textbf{Mitigated}  \\
\bottomrule
\ref{tab:techniques_VMCheck} &VM Checks &  20 & 6 \\
\ref{tab:techniques_AntiDebug} &Anti Debug &  21 & 1 \\
\ref{tab:techniques_Environment} &Resource Profiling &  6 & 6 \\
\ref{tab:techniques_TimingAttack} &Timing Attacks &  2 & 2 \\
\ref{tab:techniques_AntiDump} &Anti Dump &  2 & 0 \\
\ref{tab:techniques_CodeInjection} &Code Injections &  1 & 1 \\
\ref{tab:techniques_AntiInstrumentation} &Anti Instrumentation &  1 & 1 \\
\midrule
Total & & \totTechniques & 17 \\
\bottomrule
\end{tabularx}
\end{table}

Our current prototype can handle \totTechniques different techniques.
Table~\ref{tab:summarytechniques} shows the total number of techniques for each category introduced in Section~\ref{subsec:taxonomy}.
The majority of techniques that we cover in this work belong to either the VM Checks or the Anti-Debug categories.
This should not come as a surprise since these two categories have been documented for over a decade.
Table~\ref{tab:summarytechniques} also reports the number of techniques that \ourTool mitigates for each category.

Notice that \ourTool does not mitigate all the techniques it handles.
The rationale behind this design choice lies in the fact that most of the techniques cannot detect any component of our analysis system.
Mitigating these techniques would be not only redundant but also detrimental because each mitigation introduces overhead in the analysis.
We decided instead to mitigate only those techniques that would otherwise detect our analysis system.
These include those ``VM Checks'' that would detect our virtualized environment, those techniques that can detect Intel PIN (such as the ``Anti Instrumentation'' category), as well as all the ``Resource Profiling'' techniques which search for unusual characteristics of the software and hardware stacks (e.g., running processes, RAM and disk size).
Moreover, we mitigated both the ``Timing attack'' and ``Code Injection'' categories, which posed interesting technical challenges that we cover in more detail later in this section.

The reader can find the complete list of all the evasion techniques that \ourTool can handle inside Tables~\ref{tab:techniques_AntiDebug} to~\ref{tab:techniques_TimingAttack} in Appendix, which shows a brief description of each technique, whether it relies on APIs, CPU instructions, or memory accesses, and the mitigation strategy implemented in \ourTool (if applicable).

\subsection{Main Technical Challenges}\label{sub:MainTechnicalChallenges}

Ideally, the perfect evasion detection system would be completely transparent from the point of view of the program under analysis and able to distinguish evasion attempts from benign activities.
The design of such a system must take into consideration several aspects that, if not addressed correctly, may result in poor detection performance.

From now on, we detail the three most important challenges that we faced during the development of \ourTool and how we addressed them: time management, false-positive filtering, and code-injection detection.

Time management includes handling two different aspects: avoiding stalling attacks, aimed at timing out the sandbox, and mitigating time measurements, aimed at detecting an analysis infrastructure.
Similarly to previous solutions~\cite{polino2017measuring,d2020dissection}, our system employs a \emph{virtual-clock}. 
In addition, we also include novel countermeasures to cope with recently discovered ways to detect a virtualized environment, for instance
found in the infamous malware \emph{Locky}.
In order to reduce the number of false positives, we partition the running code into the areas containing the ``application code'' (including custom libraries and the injected one, discussed next), which we call \emph{red areas}, and the ones containing standard library code, which are considered benign.
Finally, we detect and track \emph{code injection} among threads and processes, which, to the best of our knowledge, was never covered by previous work. This issue is fundamental in the malware landscape since many malware families inject their code into other processes as a form of evasion~\cite{mitreProcessInjection}.
We tackle this issue by introducing an instrumented \emph{honeypot process}.

\noindent
\textbf{Virtual Clock}. 
To deal with stalling attacks, \ourTool uses a virtual clock, zero-initialized at the beginning of the analysis.
Time stalling techniques fall into two categories, which we name \emph{implicit} and \emph{explicit}.
A sample can explicitly require the operating system to wait for a certain amount of time, for example, by using the \texttt{Sleep} function or the underlying native API \texttt{NtDelayExecution}.
On the other hand, it may implicitly wait until a specified object is in the signaled state or until a time-out interval elapses (e.g., by using the \texttt{WaitForSingleObject} function).
In both cases, \ourTool changes the argument of such APIs to reduce the actual wait period and keeps track of how much time should have been elapsed by updating its virtual clock.
Then, when the process tries to get some information about the time, for example, through the \texttt{rdtsc} instruction or a Windows API like \texttt{GetTickCount}, we return the real-time plus the virtual clock time.
In this way, the sample is unable to distinguish whether the time has elapsed or not.
Our implementation differs from the the modified Xen hypervisor proposed by Lin et al., because the authors directly modify the clock at the hardware level, while we modify the arguments or the return value of the API.

By using \texttt{rdtsc} (the highest-resolution clock on x86 CPUs), malware can also
check if a specific snippet of code is taking an abnormal amount of time to be executed.
Such code fragments surrounded by a pair of \texttt{rdtsc} introductions are also known as ``\texttt{rdtsc} sandwiches''~\cite{oyama2019does}.
In addition, the malware authors of the Locky ransomware used a particular mechanism to detect the presence 
of a VM~\cite{lockyForcepoint} by calculating how long it takes to perform two Windows API calls, 
and comparing the results to a known ratio of at least $1/10$ (this is due to how VM products may virtualize some data structures).
This check is performed within a loop, which verifies if the ratio is matched at least once.
Simply returning a specially crafted fixed amount of time, between one \texttt{rdtsc} and another 
would not mitigate such a trick or similar ones -- as the study of Oyama et al.~\cite{oyama2019does} confirms.

\ourTool detects a \texttt{rdtsc} sandwich by checking whether two \texttt{rdtsc} occur within 50 instructions (as suggested by ~\cite{oyama2019does}).
When this happens, we calculate the $\Delta=r_2-r_1$, where $r_1$ is the result of the previous \texttt{rdtsc}, and $r_2$ would be the current one. Then, we set $r_2 = r_1 + p\cdot\Delta$ choosing, alternatively, $p=0.5$ and $p=0.05$. And, finally, we return this adjusted $r_2$.
In this way, we shrink the elapsed time by a factor of $10$ every other request.
This works because, in the case of a single \texttt{rdtsc} sandwich, it returns half of the elapsed time, while in the case of Locky's trick, the ratio is matched at least one.

We verified that our approach works by running Al-Khaser, which implements all the previously-mentioned techniques,
and verifying that none of them could detect the presence of our system. 

\noindent
\textbf{Distinguish Evasion from Other Legitimate Purposes}.
Most of the APIs, CPU instructions, and data-structures on which evasion techniques rely upon also have legitimate uses.
For instance, let us consider the example of the \texttt{cpuid} instruction again.
Windows APIs such as \emph{GetMemoryErrorHandlingCapabilities} and \emph{GetGamingDeviceModelInformation} make use, under the hood, of this instruction for legitimate purposes.
Therefore, distinguishing evasion attempts from legitimate uses might be difficult, and it requires a more in-depth understanding of the program's execution flow.

Our solution is based on two key observations: 1) real evasion attempts originate from code under control of the malware and not from third-party API calls, and 2) when evasion techniques rely on Windows APIs, they only analyze the information stored in few selected fields of the data structures that the API returns.

We leverage the first observation by establishing at runtime a \emph{red area}, i.e., a non-contiguous set of memory regions that include the memory image of the malware, any executable memory areas that it allocates, any non-standard library the malware loads at runtime, and any code that the malware injects in other processes (discussed below).
\ourTool considers the invocation of a specific API or the execution of a particular instruction as an evasion attempt only if they originate from the \emph{red area}.
If, instead, the machine instruction or the API return address belongs to another area of memory assigned to Windows libraries, \ourTool discards the event as a legitimate use.

The second observation comes into play for those suspicious APIs that initialize complex data structures, in which only a subset of fields play a role in the evasion.
For instance, this is the case for \texttt{GetSystemInfo}, which fills a \texttt{SYSTEM\_INFO} structure, whose address is provided by its caller.
In this case, only the field \linebreak \texttt{dwNumberOfProcessors} is sometimes used to evade dynamic analysis sandboxes.
Therefore, considering the access to any other field of such a structure as possible evidence of evasion would lead to many false positives. \ourTool avoids this by instrumenting the memory read operations that target only memory addresses corresponding to those specific fields. However, as soon as those addresses are written to, \ourTool ceases to monitor them
as the very same addresses may store, later during the execution, a completely different piece of data (e.g., as it is the case for stack-allocated structures).

\noindent
\textbf{Honeypot Process}.
Several malware families inject their code inside other processes' address space to evade antivirus and analysis systems~\cite{mitreProcessInjection}.
Typical targets are system processes or services that are already running when the malware starts.
The injected code is usually the one implementing the malicious activities, and, as such, the dynamic analysis should monitor its behavior.
\ourTool does so by re-routing every code injection attempt to a custom instrumented process, which we call \emph{honeypot process}, generated as soon as the analysis starts.
To this end, we instrument the execution of those Windows APIs commonly used to write into the address space (\emph{NtWriteVirtualMemory}) and start a new thread (\emph{NtCreateThreadEx}) in the context of another process.
\ourTool monitors the honeypot process the same way as the original malware process, taking care, among other things, to add its address space to the \emph{red area}.
Other solutions for dealing with code injection techniques are possible.
For example, it is technically possible to instrument the malware target processes by attaching Intel Pin to them when the injection attempts happen.
This approach, however, has two main drawbacks.
In the first place, it would risk making the entire system unstable, potentially hindering the analysis.
Moreover, extending the \emph{red area} to include the memory of an already running process would be difficult, 
as it would require to backtrack the provenance of any memory areas already allocated when the instrumentation starts.

\subsection{Instrumentation Overhead}\label{subsec:overhead}

Since our approach requires runtime instrumentation, executing a sample under \ourTool is slower than when the same sample is executed natively in the same environment.
To assess the overhead introduced by our tool, we randomly selected 1000 malware samples from our dataset among those that performed at least one ``constant'' internet connection at a certain point of their execution.
By ``constant'' we mean that the connection was easy to recognize (i.e., always contacted the same endpoint), and it was always performed simultaneously.
We then run each sample with and without our instrumentation, and we measured the time between the beginning of the execution and the flagged internet connection (as observed by an external network sniffer).
Our measures showed an average overhead of $115\pm9$\% (min 7\%, max 188\%) for the instrumented runs with respect to the baseline.

\section{Experiments Setup} \label{sec:experiments_setup}
This section presents the details of our large-scale analysis of the malware landscape to measure the prevalence and evolution of dynamic analysis evasion techniques.
We introduce the characteristics and the techniques we used to create our datasets and explain the setup of the machines we used for the experiments.
Finally, we describe how we conducted the analysis and gathered the results.

\subsection{Dataset} \label{sec:dataset}

We used four different datasets, each tailored for measuring various aspects of the phenomenon we intend to explore.
Altogether, they account for a total of \datasetsCardinality x86 PE (\emph{Portable Executable}, the standard executable file format for the Windows operating system) files.
This does not include dynamic libraries, .NET applications, nor executable installers, such as Windows Installer, InstallShield, and NSIS.
We have verified that each sample belongs exclusively to a single dataset.

\mypar{Time-based dataset.} We built this dataset to study the evolution of evasion techniques in the five years from 2016 to 2020.
We used the all-encompassing Virusshare~\cite{virusshare} dataset as a starting point, labeling each sample with the year of their first analysis on VirusTotal~\cite{virustotal}, the popular malware analysis service.
We then discarded those samples that less than ten AV flagged as malicious.
This choice is conservative compared to previous works (e.g.,~\cite{kuechler2021, cozziiotmalware} considered five detections as an indicator of maliciousness), assuring that the dataset is unlikely to contain false positives.
We also ensured that no particular malware family was over-represented by limiting each strain to at most $5\%$ of each year's total.
To this end, we relied on AVClass~\cite{sebastian2016avclass}, which labels malware samples with the most likely family based on their VirusTotal report.
This dataset accounts for 100,000 samples, equally distributed over the five years.

\mypar{Family-based dataset.} We built this dataset by collecting samples from the live feed of all files submitted to VirusTotal in a one-month timespan: from 15 December 2020 to 15 January 2021.
For each sample, we obtained its family using AVClass~\cite{sebastian2016avclass}.
If we have already collected 100 samples for such family, we discard the sample.
We ended up with 697 families for a total of 69,700 samples.
We want to emphasize that all samples were analyzed immediately after they were collected from the VirusTotal stream to minimize the chances of missing external components (c2c, website, DNS, etc.) that could affect the ability of each malware to execute correctly.

\mypar{Advanced Persistent Threat (APT).} As its name suggests, this dataset consists of malware employed by APT-groups.
We built this dataset from the two sources: the \emph{APT and Cybercriminals Campaign Collection} repository~\cite{aptgithub}, a community-maintained effort to keep track of all samples associated to targeted campaigns; the \emph{VX underground APTs}~\cite{vxapt} collection, a publicly available collection of APT-related samples from such famous malware sharing website.
Both sources are based on public malware reports of notorious companies and blog posts of security experts. 
The dataset contains \aptDatasetCardinality unique samples and presents a unique point of view,
as APT tactics are (usually) slightly different from those employed by standard malware.
In fact, besides striving to detect or evade an analysis environment directly, APT malware also 
fingerprints the environment to verify if the host is precisely the intended targeted machine.

\mypar{Goodware.} This last dataset contains a total of \goodwareDatasetCardinality non-malicious samples, which we collected from Chocolatey, a third-party package manager for the Windows platform~\cite{chocolatey}.
Our approach to collect samples consisted of installing each package indexed by Chocolatey on a vanilla Windows 10 machine.
Although its rigorous moderation review process and its strict policy on malicious and pirated software make Chocolatey a viable source for goodware, we still performed an additional filtering process that removed samples that more than two antiviruses recognized as malware (according to VirusTotal).
This filtering step discarded those benign programs with malware-like characteristics, e.g., hacking and scanning tools, which, in any case, are not representative of the average goodware.
The dataset contains manually-vetted applications, $97\%$ of which have zero detections on VirusTotal.

\subsection{Analysis Environment}

To carry out our large-scale analysis, we set up a cluster of 150 virtual machines running on a physical server powered by four Intel Xeon Platinum 8160 @ 2.10 GHz with a total of 96 cores and 512GB RAM.
We provided each VM with 2~GB of RAM and one logic CPU and installed Microsoft Windows~7. 
The filesystem was also populated with documents, spreadsheets, and other common types of files
to resemble a legitimate desktop workstation that malware may identify as a more valuable target~\cite{miramirkhani2017spotless}.

A recent study by Kuechler et al.~\cite{kuechler2021} shows the amount of code executed by malware samples tends to plateau after only two minutes, and very little additional information can be gained by running samples for more extended periods of time.
Therefore, we configured our sandbox to execute each sample for five minutes to err on the side of caution. 
Due to the overhead introduced by our instrumentation (see Section~\ref{subsec:overhead}), five minutes of runtime are roughly equivalent to two minutes and twenty seconds or real-time. 
This is also the threshold adopted by previous studies based on dynamic instrumentation~\cite{polino2017measuring}.
It is important to note that our goal is not to observe each sample's complete behavior but instead to study the techniques malware adopts to avoid dynamic analysis. 
For this reason, we expect that such techniques are concentrated in the very first seconds of execution.

%! TEX root = ../main.tex
\section{Measurements} \label{sec:measurements}

\begin{table*}[t!]
\centering
\caption{Results Over Datasets}
\label{tab:results_over_datasets}
\footnotesize
\begin{tabularx}{0.965\textwidth}{lXXXXXXXX}
        %\begin{tabularx}{0.9\textwidth}{l *{8}{>{\centering}X}}
\toprule
    &  &  & \multicolumn{5}{c}{\textbf{Yearly Datasets}}  & \\
    \cmidrule(r){4-8}
    & \textbf{Goodware} & \textbf{APT} & \textbf{2016} & \textbf{2017} & \textbf{2018} & \textbf{2019} & \textbf{2020} & \textbf{Families} \\
    % & \textbf{Goodware} & \textbf{APT} & \textbf{2016} & \textbf{2017} & \textbf{2018} & \textbf{2019} & \textbf{2020} & \textbf{Families} \\
\toprule
%\textbf{Dataset Cardinality}        &  \goodwareDatasetCardinality & \aptDatasetCardinality & 20,000 & 20,000 & 20,000 & 20,000 & 20,000 & 110,000 \\
%\toprule
\textbf{Started} & 4,905    & 3,601    & 17,094   & 18,367   & 18,852  & 17,173   & 18,306   & 59,468   \\
%                 & (33.9\%) & (95.5\%) & (85.5\%) & (91.8\%) &(94.4\%) & (85.9\%) & (91.5\%) & (88.5\%) \\
\midrule
\textbf{Active}                 & 52.0\% & 96.5\% & 99.1\% & 99.3\% & 99.1\% & 98.1\% & 98.4\% & 95.2\% \\ 
\midrule
\textbf{Evasive}                &  8.4\% & 29.6\% & 32.0\% & 32.0\% & 29.8\% & 35.8\% & 40.4\% & 42.1\% \\ 
\midrule
\textbf{Active \& Evasive}      &  8.0\% & 29.6\% & 32.0\% & 32.0\% & 29.6\% & 35.7\% & 40.3\% & 41.7\% \\ 
\midrule
\textbf{AVG N. of Techniques}   & 1.3$\pm$0.9 & 1.7$\pm$1.5 & 1.7$\pm$1.8 & 1.9$\pm$1.8 & 1.9$\pm$1.6 & 1.8$\pm$1.6 & 2.1$\pm$1.9 & 1.6$\pm$1.5 \\
\midrule
\textbf{MAX N. of Techniques}   & 12 & 13 & 13 & 13 & 13 & 14 & 14 & 34 \\
\midrule
\textbf{Internet Connection}    &  0.5\% & 11.2\% & 31.0\% & 37.4\% & 30.6\% & 15.5\% & 13.2\% & 12.4\% \\ 
\midrule
\textbf{Child Process}          &  6.2\% & 43.9\% & 38.7\% & 50.0\% & 42.7\% & 48.5\% & 39.4\% & 40.7\% \\ 
\bottomrule
\end{tabularx}
\end{table*}

This section presents the analysis results we performed over a dataset of \datasetsCardinality Windows PE samples.
In this analysis, we consider that a sample has \emph{started} if it invoked at least one native API, while we consider it \emph{active} if it executed at least 50 native API invocations -- we adopted the same threshold of Kuechler et al.~\cite{kuechler2021}.
Before presenting our results, we discuss how False Positives (FP) and Negatives (FN) could affect our measurement.

\subsection{False Negatives}

To assess that our implementations of the detection and mitigation strategies are sound, we conducted two experiments to uncover possible false-negative results, i.e., known evasive techniques that \ourTool failed to detect.

For the first experiment, we used \ourTool to analyze open-source tools that implement large numbers of evasion techniques.
More in detail, we examined binaries built from Pafish~\cite{pafish}, SEMS~\cite{SEMS}, VMDE~\cite{VMDE}, and Al-Khaser~\cite{alkhaser}.
Our system correctly identified \emph{all} the evasive tricks performed by these tools.

In the second experiment, we analyzed publicly documented evasive malware samples.
In particular, we collected a curated list of 37 samples, of which:
15 were labeled as evasive by JoeSecurity~\cite{joeEvasiveReports} reports, and 22 were chosen according to online forums
in which researchers performed in-depth analysis and identified the presence of evasive techniques.
By manually reverse-engineering each sample, we confirmed the presence and nature of each evasive strategies.
The subsequent analysis performed by using \ourTool successfully detected
and mitigated all evasion techniques we confirmed during the manual
analysis stage.

\subsection{False Positives}

While malware authors can use evasive techniques to detect the presence of a dynamic analysis environment, some of these techniques can also have legitimate use cases that may not constitute evidence of evasion attempts.
For instance, an application might need to request the cursor position to open a pop-up menu or retrieve the number of CPUs to better distribute intensive computation among different threads.

A possible way to detect these false alerts could be to execute a sample twice for each detected technique: the first time with the corresponding mitigation enabled and the second without it. 
A completely different behavior or an abrupt termination of the execution when the mitigation is disabled could be a sign that the sample was indeed employing for evasion.

Unfortunately, this solution does not apply to all the techniques for three reasons: I) while we can detect all techniques, not all of them have corresponding remediations; II) in case of stalling-like timing attacks, disabling the remediation would require running the sample for long periods of time, making this approach too time-consuming; III) in some cases, we cannot predict which exact data the program uses to reveal the presence of the analysis environment (e.g., the process enumeration technique --- from now on \texttt{process\_enum} --- could be used to detect suspicious parent process names or to find antiviruses running on the system, both of which could deter the malware from starting).

Therefore, we tried to estimate the number of FP by using a data-driven approach.
In particular, we focused on the top 10 evasive techniques among those for which we provide remediation or those we can craft ad-hoc data to ``fool'' the running program.
For example, we disabled the remediation to the \texttt{RDTSC} by disabling the virtual clock, and we changed the return value of \texttt{IsDebuggerPresentAPI} to make the process believe that it is running under a debugger (when in fact it is not, since our approach uses DBI).

We randomly selected $10,000$ \emph{active} samples -- distributed among different families -- among those that adopted such evasive techniques. 
We then executed each sample twice, the first time while enabling the remediation for the target technique, the second time without the remediation, or while providing fake ad-hoc data.
Then, for each sample, we compared the results of the two executions, checking if they triggered the same techniques \emph{and} the same APIs with externally visible effects (i.e., those that modify the environment by writing to files, updating registry keys, or establishing network connections), reporting the corresponding technique as FP if that was not the case.
Below we discuss the top four techniques separately and then the remaining six because we came to different conclusions on these two groups.

\mypar{False Positives estimate - Top 4.}
The four most common evasion techniques in all our datasets, listed in descending order of prevalence, are:
\begin{enumerate}[noitemsep,topsep=2pt,parsep=2pt,partopsep=2pt]
\item \texttt{GetTickCount}, which uses the homonymous Windows API function to retrieve the number of milliseconds since the last reboot. 
\ourTool modifies the result as described in Section~\ref{sub:MainTechnicalChallenges}.
\item \texttt{cpuid\_is\_hypervisor}, which checks the presence of an hypervisor by executing the instruction \texttt{cpuid} (with \texttt{EAX=1}), and then checking whether the $31^{\mathrm{th}}$ bit of \texttt{ECX} register is set. \ourTool always resets that bit.
\item \texttt{GetCursorPos}, a Windows API that retrieves the position of the mouse cursor in screen coordinates. Malware uses this API to find evidence of sandboxes, in which the mouse cursor is still for a long time. \ourTool mitigates this technique by returning different coordinates at each call.
\item \texttt{NumberOfProcessors}, named after a field in the Process Environment Block, is often used by malware to identify a sandbox environment with only one CPU core. When a sample access this field, \ourTool always sets it to four. 
\end{enumerate}

In each dataset, (1) occurs in more than the 55\% of samples, (2) in more than 50\%, (3), and (4) in more than 20\%.
For these four techniques, our FP tests reported that in $92\pm1\%$ of cases, the samples exhibited the same behavior with or without the mitigation. 
In other words, in the vast majority of cases, the samples did not exit after performing those potentially evasive actions, nor they significantly deviated from their original behavior.

To understand \emph{why} these four techniques are so prevalent for non-evasive purposes, we manually analyzed some of goodware programs that used them. 
We found that \texttt{GetTickCount} is often used to estimate the remaining time of potentially long operations or to check whether the user is idle (this can be useful to optimize rendering in text editors, for instance). 
\texttt{GetCursorPos} is often used to show pop-up menus near the mouse cursor and handle drag-and-drop operations.
While these are somewhat expected usages, we also noticed some curious use cases.
For instance, some programs use these two functions as sources of entropy to initialize pseudorandom number generators.
The \emph{clang} and \emph{MinGw} C compilers also use \texttt{GetTickCount} in the initialization routine of the \emph{stack canary},
and the Microsoft Visual C++ compiler inserts the \emph{cpuid} instruction as part of the program initialization routine.
Even after excluding common compiler-generated patterns, the usage of the \texttt{cpuid} instruction remained very common among all samples,
as it can be used to retrieve the number of processors for allocating thread pools or check for hardware-assisted encryption.

In conclusion, while a small fraction of malware used these four techniques for evasion (particularly the \texttt{cpuid} one), their general and legitimate use is too widespread and would pollute our findings with too many false positives. 
Therefore, we excluded these techniques from the rest of the discussion.

\mypar{False Positives estimate - from 5 to 10.}
The remaining six techniques we tested in our FP evaluation are, in descending order of prevalence: \texttt{IsDebuggerPresentAPI}, \texttt{disk\_size\_getdiskfreespace}, \texttt{RDTSC}, \texttt{IsDebuggerPresentPEB}, \texttt{memory\_space}, and \texttt{NQIP\_Process\_debug\_flag}.

For the three anti-debug related techniques, samples exhibited the same behavior just in $4\%$ of the cases, while we obtained a similar result with \texttt{RDTSC} for $6\%$ of the samples.
Finally, the two Resource Profiling techniques \texttt{disk\_size\_getdiskfreespace} and \texttt{memory\_space} (which we mitigated by returning small values: 5GB of free disk space and 1 GB of installed RAM), resulted in samples showing the same behavior respectively in $33\%$ and $26\%$ of the cases.

Our investigation shows that four techniques are very frequently used without evasive purposes (and thus we removed them from the rest of our analysis), four are commonly used for evasion, and the two Resource Profiling techniques are somehow in a grey area.
While their use is not always clearly related to evasion, we decided to report them in the rest of the paper because, in around 70\% of the cases, they made the behavior of a sample diverge.
Therefore, their prevalence has a clear impact on the data collected by a sandbox and could help other researchers to investigate their use further.

As the last point, it is worth noting that, during a preliminary analysis, we found programs that rewrite their in-memory PE headers with the same values already present; that is, without actually changing anything.
For this reason, \ourTool not only intercepts memory writes to those areas but also checks the written values to determine that a change happened in order to make a sound detection of \texttt{ErasePEHeader}.
To give a rough idea, we found this \emph{potential} false-positive among samples accessing the PE header in 95\% of goodware (in an initialization routine) and less than 1\% of malware.

\noindent \fbox{\begin{minipage}[c]{0.96\columnwidth}
\textbf{Takeaways}: We found four techniques having more legitimate uses than evasive; therefore, future works must take this into account, and existing tools that detect evasive techniques should remove or notify their users about those potential false positives. 
In contrast, Resource Profiling techniques that might seem harmless, like checking for free space on the hard drive or in RAM, may hide a malicious attempt to evade analysis.
\end{minipage}}

\subsection{General Statistics}

Table~\ref{tab:results_over_datasets} provides some general statistics about our measurements on the different datasets.
All percentages refer to the number of \emph{started} samples, which account for 85\% of total malware and 50\% of the benign samples.

The relatively high share of benign applications that failed to start was due to missing dynamically linked libraries.
Indeed, even if standalone programs exist, ordinary Windows applications require to be installed by completing a procedure that stores all program dependencies and sets up the needed registry keys.
However, we do not consider this a limitation as the de-facto standard scenario in malware analysis involves collecting and analyzing \emph{executables}, often submitted by a third party for dynamic analysis. 

\mypar{Evasive Samples.} We consider a sample \emph{evasive} if it employed at least one evasive technique (apart from those four mentioned above) during its runtime.
Even if our malware datasets differ in cardinality and typology, the percentage of evasive behavior always lies between 30-40\%, while it drops drastically to 8\% for goodware.
The average number of techniques per evasive sample (fifth row in Table~\ref{tab:results_over_datasets}), and the fact that their median is $1$ in each dataset, suggest that each malware sample tends to rely only on a few techniques.

The fact that the percentage of both active \emph{and} evasive samples matches the rate of evasive samples across all datasets suggests that our mitigations were successful in deceiving malware. 
Otherwise, evasive malware would likely terminate after detecting the analysis environment, and therefore it would fail to count as an active sample.

To validate the hypothesis that malware exhibits their malicious behavior only after their evasion techniques, we counted the number of API calls with ``externally visible effects'' (described previously) that each sample invoked before employing the first evasion techniques, as well as those invoked after the last evasion techniques.
For 73\% of the malware, 10\% of such API calls appear before the first technique, while 77\% after the last one.
Nearly three-quarters of malware has precisely the behavior we expected: no or little activity before the first evasive technique and a much more intense activity after the last evasive technique.

\begin{table}[t!]
    \caption{Most used Evasive Techniques}
    \label{tab:techniques_Occurrencies}
    \footnotesize
    \begin{tabular}{c c c}
    \textbf{Malware [2016, 2020]} & \textbf{APT} \\
    \bottomrule
    \texttt{ErasePEHeader}  & \texttt{IsDebuggerPresentAPI}  \\
    (8.9\%) & (7.5\%) \\
    \midrule
    %\texttt{IsDebuggerPresentAPI} & \texttt{disk\_size\_getdiskfreespace}  \\
    \texttt{IsDebuggerPresentAPI} & \texttt{disk\_getdiskfreespace}  \\
    (7.1\%) &(5.2\%) \\
    \midrule
    \texttt{RDTSC} & \texttt{RDTSC} \\
    (5.9\%) & (4.1\%) \\
    \midrule
    \texttt{process\_enum}  & \texttt{CanOpenCsrss} \\
    (5.2\%) & (3.0\%) \\
    \midrule
    \texttt{vm\_check\_mac} & \texttt{process\_enum} \\
    (4.1\%) & (2.7\%) \\
    \midrule
    \texttt{IsDebuggerPresentPEB} & \texttt{IsDebuggerPresentPEB} \\
    (4.1\%) & (2.5\%) \\
     \midrule
     \texttt{disk\_getdiskfreespace} & \texttt{time\_stalling} \\
     %\texttt{disk\_size\_getdiskfreespace} & \texttt{time\_stalling} \\
    (3.8\%) & (2.3\%) \\
    \midrule
    \texttt{SizeOfImage} & \texttt{SizeOfImage} \\
    (3.8\%) & (2.2\%) \\
    \midrule
    \texttt{memory\_space } & \texttt{vm\_check\_mac} \\
    (2.4\%) & (2.1\%) \\
    \midrule
    \texttt{CanOpenCsrss} & \texttt{NSIT\_ThreadHideFromDebugger} \\
    (2.3\%) & (1.4\%) \\
    \bottomrule
\end{tabular}
\end{table}

\mypar{Techniques occurrence.} The left column of Table~\ref{tab:techniques_Occurrencies} shows the most common techniques that \ourTool detected, ordered by frequency, for the samples in our time-based dataset.
Among the most frequently encountered techniques, the majority belong to either the Anti Debug or Resource Profiling categories.
Overall, the \texttt{RDTSC} technique is the third most used, making it the de-facto standard for Timing Attacks.

Table~\ref{tab:techniques_Occurrencies} also describes the differences in terms of techniques occurrences between ``regular'' and APT malware.
One can notice that APT samples do not seem interested in removing their PE headers from memory, which is instead the most common evasive technique among traditional malware samples.
On the other hand, the presence of \texttt{time\_stalling} in the second column shows that APT are more focused on timing attacks.

\noindent \fbox{\begin{minipage}[c]{0.96\columnwidth}
\textbf{Takeaways}: 30-40\% of malware use at least one evasive technique. Evasive malware use, on average, few techniques. Anti Debug and Resource Profiling are the most common form of evasion. The \texttt{RDTSC} technique/instruction is the third most frequent, indicating a great interest on the part of malware authors to find out whether time has been accelerated or there is some component that slows it down. APT samples employ Anti Dump techniques significantly less often, and they are more focused on timing attacks.
\end{minipage}}

\subsection{Occurrence of the evasive behavior}

\begin{figure*}[t]
\includegraphics[width=\textwidth,height=5cm]{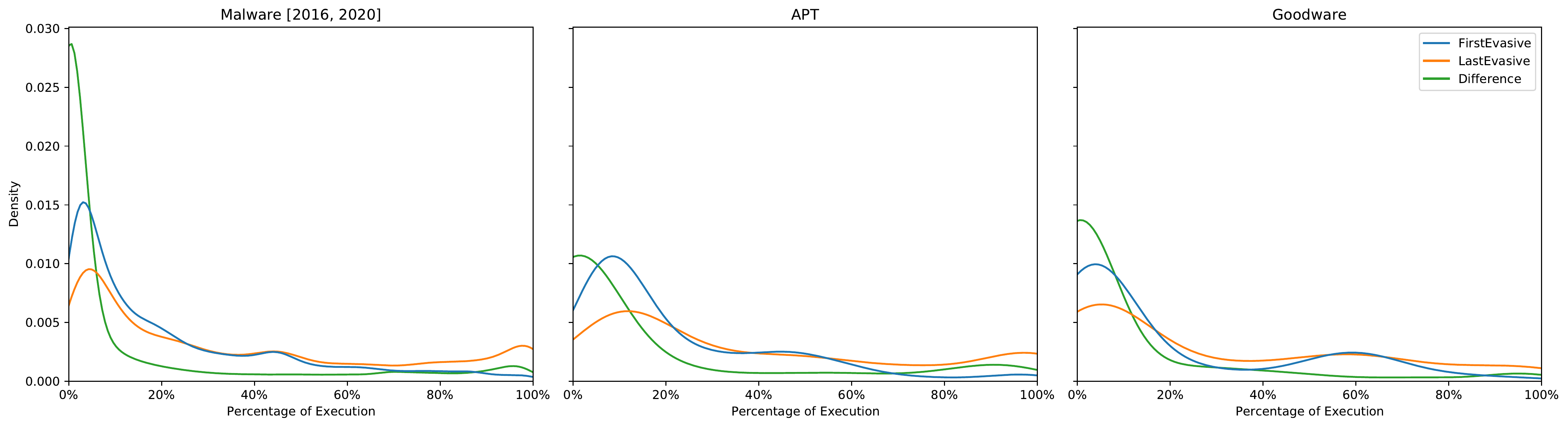}
\caption{Distribution of First and Last evasive technique, and their difference, during normalized execution time}
\label{fig:FirstLastDiff}
\end{figure*}

Intuitively, one would expect evasive malware to check whether it is under analysis at the beginning of its execution.
To test this assumption, we normalized the execution time of each sample in a $[0,100]$ range, and then we tracked when the first and last evasive techniques were performed; we also consider the difference between the last and the first to see whether they are usually executed in quick succession or at different points in time.
The kernel density estimate plots in Figure~\ref{fig:FirstLastDiff} show the distribution of observations respectively for malware, goodware, and APT samples.
The probability of the first and last evasion techniques occurring during the first 10\% of the execution are respectively 47\% and 34\% (for malware), 43\% and 24\% (for APT), and 64\% and 49\% (for goodware).

From the plots it is clear that the evasive techniques are performed at the beginning of the process, especially for goodware.
However, there is a marked difference between common malware and APT; the former tend to use the techniques next to each other in quick succession, the latter uses them in a more distributed fashion during the execution.
Despite being a minority, the graphs also show that a non-negligible amount of samples use evasive techniques throughout their entire execution. 
For instance, 11\% of the malware samples still employed some evasive techniques during the \emph{last} 10\% of their execution (versus $17\%$ of APTs and $8\%$ of goodware).

We also investigated which are the three most common categories that are used at the beginning, in the middle and at the end of the execution. 
Therefore, we divided the normalized execution time in three slots: [0-10], [11-89], and [90-100]. 
For each slot, we computed the percentage of how many times a certain category has been used first. 
All the results are shown in Table~\ref{tab:top3firstrange} in the Appendix due to space limit, although we report the relevant observations. 
APT and malware use the same categories at the beginning: Anti Debug, VM Checks, and Timing Attacks. 
They have respectively less than the 71\%, 12\%, and 5\% share in both datasets.
Then, Anti Debug drops to third position ($>$ 15\%) at the middle, and in fourth position ($>$ 7\%) at the end. 
This fact clearly shows that malware authors are more interested in detecting debuggers at the beginning of the execution code.

\noindent \fbox{\begin{minipage}[c]{0.95\columnwidth}
\textbf{Takeaways}: Evasive behavior occurs mainly at the beginning of the process, but at least 10\% of the malware continues to check throughout the execution. Malware authors focus on detecting debuggers at the beginning of the execution. 
\end{minipage}}

\subsection{Order of appearance}
The order in which different techniques are used by malware authors is an important factor, 
because if researchers fail to properly mitigate them, the results would be skewed towards those evasive
techniques that are used first (as the others would not be observed at all).

Whenever a malware sample employed two or more categories of anti-analysis
techniques, AntiDebug was used first in 88\% of the cases. Within the APT
dataset, the percentage increased to a stunning 95\%.
It is difficult to tell why AntiDebug is almost unanimously chosen 
as first line of defense by malware authors, but a possible explanation is 
that it is used to prevent human analyst to overcome other protections (such as 
packing and VM-detection).

In the remaining cases in which AntiDebugging was not first, the first technique employed by malware was a form of Timing
Attacks (44\%), VM Checks (38\%), or Anti Dump (15\%). APT samples used
instead first Profiling (38\%), Timing Attacks (33\%), or VM Checks (27\%). 
In this context and for both datasets, the most common technique for Timing
Attacks is the use of \texttt{RDTSC} instruction, while for VM Checks is
\texttt{vm\_check\_mac}.

Our hypothesis for the prevalence of Resource Profiling in the APT is that
it is not the type of threat that usually tries to infect as many machines
as possible, but it is very targeted to hit a particular objective.
Therefore, it is likely that they are collecting as much information as
possible to check if they are running on the right target.

\noindent \fbox{\begin{minipage}[c]{0.95\columnwidth}
  \textbf{Takeaways}: Anti Debug is the category of techniques that almost always appears first. Moreover, malware analysis sandboxes should 
  counteract at least the use of the \texttt{RDTSC} instruction to control the passage of time.
  \end{minipage}}

\subsection{Yearly Evolution}

\begin{figure}[t]
    \centering
    \includegraphics[width=0.6\columnwidth]{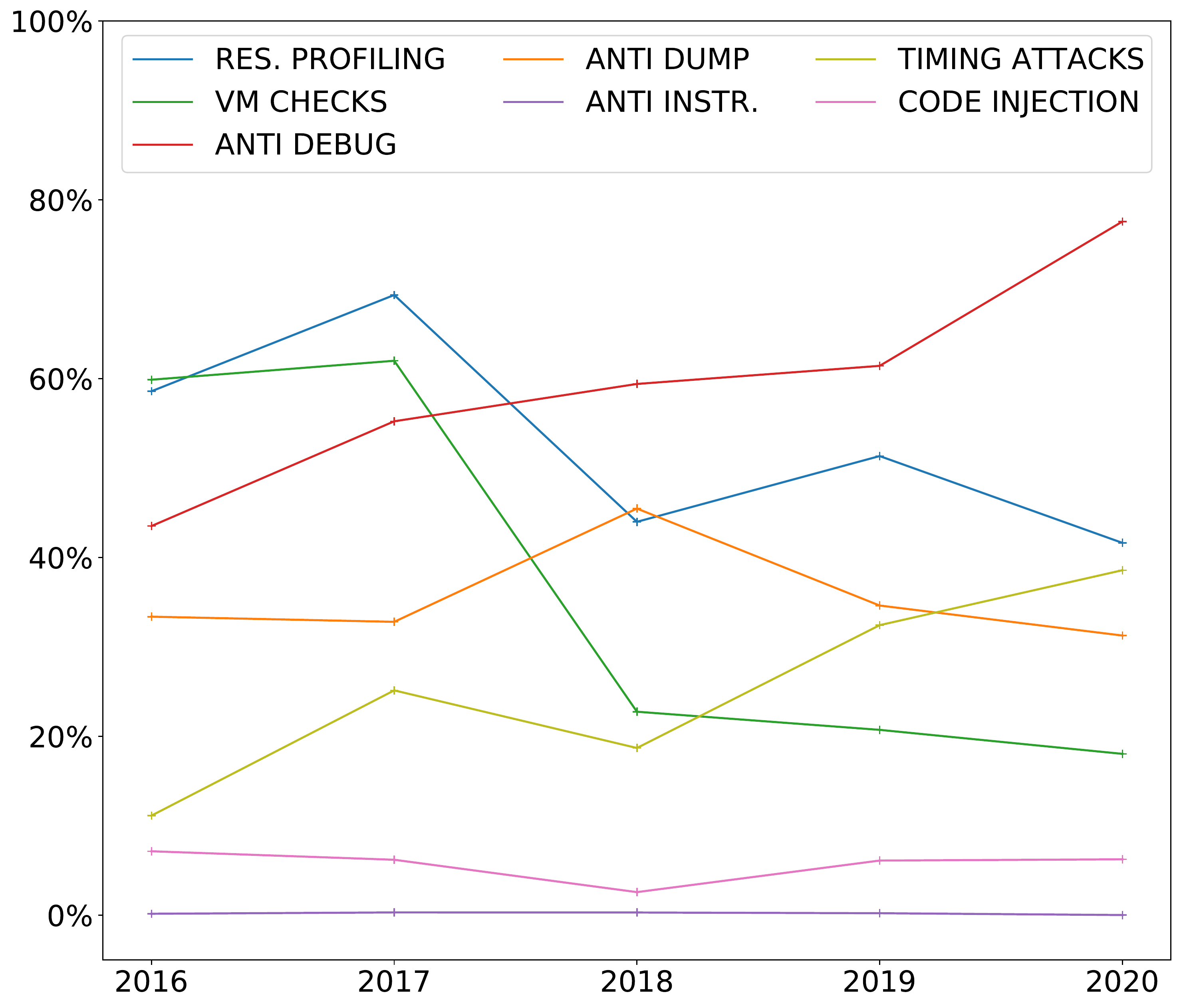}
    \caption{Evasive category usage over the years}
    \label{fig:lineplot_evasion_trends}
\end{figure}

We now look at the evolution over the years on the use of different categories (see Section~\ref{subsec:taxonomy}) and individual techniques.
Figure~\ref{fig:lineplot_evasion_trends} shows the evolution of the incidence of each category.
We can identify three types of trends: increasing, steady, and decreasing.
The first trend is characteristic of the Anti Debug and Timing Attack categories.
The former has been growing steadily over the years, and it is now clearly the most common type of evasive technique employed by malware.
The most used techniques in this category are \emph{IsDebuggerPresentPEB} and \emph{IsDebuggerPresentAPI,} which respectively check the value of the \emph{BeingDebugged} field at PEB structure and use the \emph{IsDebuggerPresent} API to detect a debugging environment.
Since both are relatively easy to address, this is somehow a good news for malware analysts.
The incidence of Anti Dump, Anti Instrumentation, and Code Injection remained stable during the considered period.
However, while the first hold around 40\%, the other two categories were much rarer, with Anti Instrumentation being the least prevalent technique every year, averaging at 0.4\%.

We can also observe a decreasing trend for VM Checks and Resource Profiling.
For what concerns the former, \texttt{vbox\_check\_mac} -- which detects virtualized environments by inspecting the manufacturer field of the MAC address -- was the most common technique in both 2016 and 2017, before falling in eighth place in 2018 and disappearing from the top 20 in 2020.
A possible explanation for the malware's abandonment of this technique (and VM Checks in general) could be that server virtualization (usually by type-1 hypervisors) has become common in the enterprise world.
Not infecting virtualized machines would be counterproductive for malicious actors, and the benefits of evading virtualized dynamic analysis may not justify this choice anymore.
Regarding Resource Profiling, the most frequent resources that malware profiles for evasion used to be the remaining disk space and RAM size, respectively.
The first was in the top 6 until 2019, before dropping to 16th place in 2020.
After 2018, only \texttt{process\_enum} seems to remain prevalent among the Resource Profiling category, while the adoption of others, which were very common until 2019, drastically decreased.

We also found four techniques that made their first appearance in 2017 (\texttt{SharedUserData\_KernelDebugger} and \texttt{Firmware\_RSMB}) and 2018 (\texttt{MemoryBreakpoints\_PageGuard} and \texttt{Firmware\_ACPI}).
The interested reader will find the plots with each year's most common evasion techniques in Appendix (Figure~\ref{fig:techniques_prevalence_oty}).

\noindent \fbox{\begin{minipage}[c]{0.95\columnwidth}
    \textbf{Takeaways}: As virtualized environments become more and more attractive for malware, VM Checks techniques decrease in popularity to the point of not being worth mitigating. On the other hand, Anti Debug techniques are now the most commonly adopted techniques.
\end{minipage}}

\subsection{Evasion among Malware Families}
\label{subssh :families}

In our analysis of the results produced on the family-based dataset (see Section~\ref{sec:dataset}), we found that $472$ out of $679$ (68\%) families contain at least one evasive sample, i.e., a sample that performs at least one evasion technique.
Surprisingly, only 33 of the analyzed families, including the infamous \emph{clop} ransomware, employed evasive techniques in \emph{all} their samples.

Another metric to assess a malware family's overall evasiveness is by counting the number of evasion techniques that each sample in the family adopts.
According to this metric, the three most evasive families in our dataset are \emph{cnbtech}, \emph{dothetuk}, and \emph{chepro,} for which the median value of the number of evasion techniques per sample is $10$, $10$, and $9$, respectively. 
We also found a family with a tremendous number of techniques: while in the other datasets, the maximum number of techniques employed by a sample ranged between 12 and 14, the most evasive samples of the \emph{mwpm} family adopted 34 evasive techniques.

Since samples of the same malware family often share behavioral traits, it is reasonable to assume they could also employ the same evasive techniques.
To validate this hypothesis, we performed measurements to evaluate the similarities of the evasion techniques employed by the evasive samples of each family.
In particular, we begin by defining, for each sample, the set of evasion techniques that it uses.
For each family, we then calculate the set intersection of its samples' evasion techniques; from now on, the \emph{evasive footprint}.
For $249 (53\%)$ families out of the 472 that have at least one evasive sample, the evasive footprint was not empty.
This means that there was at least one technique that \emph{all} the samples in such a family share.
The family with the highest number of common evasion techniques was \emph{dothetuk} with $8$, followed by \emph{cnbtech} (7) and \emph{asyncrat} (4).

\noindent \fbox{\begin{minipage}[c]{0.95\columnwidth}
\textbf{Takeaways}: Within the same family, it is very likely (95\%) to encounter samples both using and not using evasive techniques.
In about half of the families, when a sample uses an evasive technique, all the other samples share such technique.
\end{minipage}}

%! TEX root = ../main.tex

%! TEX root = ../main.tex
\begin{table*}
\centering
\caption{Packers and Protectors over the datasets}
\label{tab:packers_protectors}
% \fontsize{6.0}{9}\selectfont
\small
\begin{tabular}{ccccccccc}
\toprule
& \textbf{APT} & \textbf{2016} & \textbf{2017} & \textbf{2018} & \textbf{2019} & \textbf{2020} & \textbf{Families} \\
\toprule
\textbf{$Started$}       & 3,601    & 17,094   & 18,367   & 18,852  & 17,173   & 18,306   & 59,468   \\
\midrule
$$\mbox{\footnotesize$\frac{Packed}{Started}$}$$ & 7.6\%  & 3.1\% & 5.8\% & 7.0\% & 8.9\% & 6.2\% & 24.5    \% \\ 
\midrule
$$\mbox{\footnotesize$\frac{Evasive}{Packed}$}$$ & 82.9\%  & 70.5\% & 62.1\% & 62.6\% & 60.7\% & 79.0\% & 61.4\% \\ 
\midrule
$$\mbox{\footnotesize$\frac{Protected}{Started}$}$$ & 3.9\%  & 0.9\%  & 1.4\%  & 2.1\%  & 2.6\%  & 2.8\%  & 3.9\% \\ 
\midrule
$$\mbox{\footnotesize$\frac{Evasive}{Protected}$}$$ & 53.1\% & 57.7\% & 64.1\% & 65.3\% & 57.8\% & 55.2\% & 86.3\% \\ 
\bottomrule
\end{tabular}
\end{table*}

\section{Evasion w.r.t. Packers \& Protectors}

Our work also investigated the correlation between packing, software protectors, and evasive behaviors in malware.
We used \emph{Detect It Easy}~\cite{detectiteasy} (DIE) to determine whether each sample in our dataset uses either packing or protection techniques.
As Mantovani et al.~\cite{mantovani2020prevalence} showed, DIE performed significantly better than other signature-based packer/protector detectors.

It is worth noting that our goodware dataset is not representative of the prevalence of packing and software protectors.
In fact, it under-represents commercial applications like videogames, which is the software class that employs software protection mechanisms more often.
Developers of freeware and open-source applications (which our dataset primarily represents) are less likely to employ packing or software protection.
It is not surprising that both techniques combined account for less than the $0.6\%$ of goodware samples. 
Therefore, in the remainder of this Section, we focus on malicious samples.

For simplicity, we use the term ``packed'' for samples that DIE recognizes as compressed with an off-the-shelf packer, 
and ``protected'' for samples that DIE recognizes as obfuscated using known software protection programs.
Table~\ref{tab:packers_protectors} summarizes the results of the prevalence of packing and software protection among our datasets.
The second and fourth rows represent the percentage between the number of packed and protected samples, respectively, and the number of started samples (as defined in Section~\ref{sec:measurements}).
Instead, the third and fifth rows show the ratio between the number of samples that exhibited at least one evasive behavior and those (started) that are packed or protected, respectively.

It is worth emphasizing again that we only report off-the-shelf packers and protectors that DIE recognizes with a static signature; 
thus, that the data has a clear bias towards popular solutions and fails to consider samples that use custom packing techniques.

\mypar{Packers}
UPX and ASPack are the most prevalent packers in our malicious datasets and account for $75\%$ and $12\%$ of the packed samples, respectively, followed by PECompact ($7\%$) and NsPacK ($1\%$); all others account for less than $1\%$.
Even if evasive techniques are typical of protectors, they are more common among packers.
Indeed, more than $60\%$ of packed samples exhibited evasive traits.
In particular, we found evidence of Anti Dump techniques in more than $90\%$ of packed samples in all datasets
--- which is a natural choice to prevent an analyst from dumping the unpacked code from memory.

\mypar{Protectors}
We found evidence of evasive behaviors in more than $50\%$ of protected samples, regardless of the dataset.
While packed samples rely mostly on Anti Dump techniques, protected samples employ many different classes.
On average, each protected evasive sample uses $4.1\pm2.3$ distinct techniques, almost twice as those of other evasive samples considered in all datasets.
Overall, DIE reported 111 different protectors in our datasets, but the ones with at least a 10\% share are, in decreasing order: \emph{Themida}, \emph{VMProtect}, \emph{ASProtect}, and \emph{ENIGMA}.
Themida is by far the most widespread, to the point that it accounts for more samples than the other three most common protectors combined (probably because the free version has more features and fewer limitations than the competition).
For this reason, we studied if we could mitigate the techniques of this protector by downloading its demo version and using it to protect a simple ``Hello World'' program.
In doing so, we made sure to enable every evasion techniques that Themida provides and then
verified that \ourTool was able to detect them all and force the program to ran correctly.

\noindent \fbox{\begin{minipage}[c]{0.95\columnwidth}
\textbf{Takeaways}: $90\%$ of samples packed with off-the-shelf packers overwrite their PE header in memory. 
Themida is the most common protector and also the one with the most evasive behavior. 
In general, samples that use off-the-shelf protectors are the ones most likely to employ evasive techniques.
\end{minipage}}

%! TEX root = ../main.tex
\section{Conclusions} \label{sec:conclusions}
This paper presents the first large-scale analysis on the prevalence of anti dynamic analysis evasion techniques in Windows malware.
We developed our own tool to conduct the experiments.
While it was largely inspired by both \emph{Arancino} and \emph{Bluepill}, 
our system 
implements a more fine-grained detection mechanism
and covers a much broader range of techniques and categories.

Our measurement includes $180K$ samples from four diverse datasets:
malicious samples observed over the last five years, recent samples grouped
by family, APTs, and goodware. Our analysis achieved an activation rate
above $95\%$ for malware samples. We cannot tell whether the remaining 5\%
stopped their execution because they detected the analysis environment by
using some undocumented technique or, more likely, because they missed some
dependencies or because their online infrastructure was not reachable
anymore. This did not prevent us from capturing the most
precise picture to date about the current prevalence of evasion behaviors.

Circumventing evasion techniques is a cat-and-mouse game, and we are aware that malware authors can target and evade \ourTool in the future.
However, the goal of this work was to study the \emph{current} picture.
Our data lead to numerous findings and interesting trends: we witnessed the rise of anti-debug and timing-attack
and a surprisingly low number of code-injections over the last five years.
Anti-Debugging is almost always used first, and while 
evasion techniques are more prevalent in the first part of 
execution, some samples continue to mix evasive and ``normal'' behavior for their entire lifetime.

While we anticipated some of the results, others were surprising. For instance, we discovered that malware authors are more likely to protect against debuggers than VMs and emulators and that APT and targeted samples are not more ``protected'' than general malware. 
We also discovered, against common sense, that the number of different anti-analysis techniques employed by each sample is low (e.g., in the ``Yearly Datasets,'' only 41\% of malware uses more than one and 17\% more than two).
This quantification provides the building blocks for any data-driven solution fighting evasive Windows malware and is essential for everyone working in the malware analysis field. 
For instance, knowing which percentage of samples would not run in a VM is paramount to the design of an analysis infrastructure, and knowing the fraction of samples that perform process injection can help assess the impact of sandboxes that cannot track this activity.

\newpage
\bibliographystyle{IEEEtran.bst}
\bibliography{biblio.bib}
%! TEX root = ../main.tex
%\onecolumn
\section{Appendix}
\label{appendix}

%! TEX root = ../main.tex
% \usepackage{multirow}

\begin{table}[h]
    \centering
    \caption{Top Three Evasive Categories First Measured w.r.t. Execution Range}
    \label{tab:top3firstrange}
    \begin{tabular}{|c|l|c|l|c|l|c|} 
    \cline{2-7}
    \multicolumn{1}{l|}{} & \textbf{[0-10]} & \textbf{\%} & \textbf{[11-89]} & \textbf{\%} & \textbf{[90-100]} & \textbf{\%} \\ 
    \hline \hline
    \multirow{3}{*}{\textbf{Mal}}  & Anti Debug          & 74 & Anti Dump           &  45 & Resource Prof.  & 44 \\ 
    \cline{2-7}
                                       & VM Checks           & 12 & Timing Attacks      &  20 & Anti Dump           & 32 \\ 
    \cline{2-7}
                                       & Timing Attacks      &  5 & Anti Debug          &  18 & Timing Attacks       & 17 \\ 
    \hline \hline
    
    \multirow{3}{*}{\textbf{APT }}     & Anti Debug          & 71 & VM Checks           & 51 & Timing Attacks      &  37 \\ 
    \cline{2-7}
                                       & Resource Prof.  & 16 & Timing Attacks      & 31 & Resource Prof.  &  32 \\ 
    \cline{2-7}
                                       & Timing Attacks      & 10 & Anti Debug          & 15 & VM Checks           &  10 \\ 
    \hline \hline
    \multirow{3}{*}{\textbf{Good}} & Anti Debug          & 54 & Anti Debug          & 41 & Resource Prof.  & 46 \\ 
    \cline{2-7}
                                       & Resource Prof.  & 28 & Resource Prof.  & 38 & Anti Debug          & 34 \\ 
    \cline{2-7}
                                       & Timing Attacks      & 17 & Timing Attacks      & 20 & Timing Attacks      & 19 \\
    \hline
    \end{tabular}
\end{table}

\begin{figure*}[]
    \centering
        \subfloat[2016]{\includegraphics[width=0.6\columnwidth]{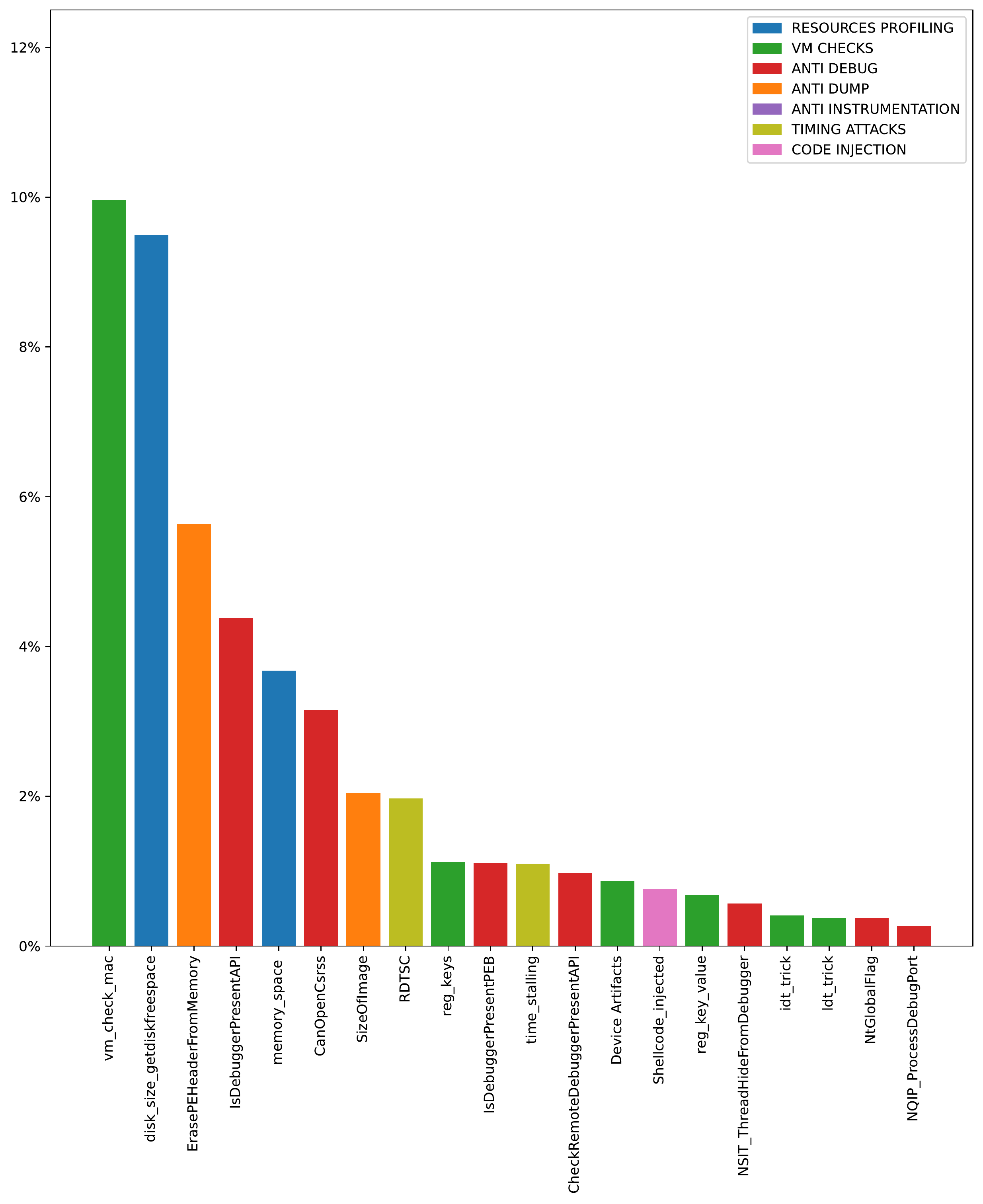}}\hfil
        \subfloat[2017]{\includegraphics[width=0.6\columnwidth]{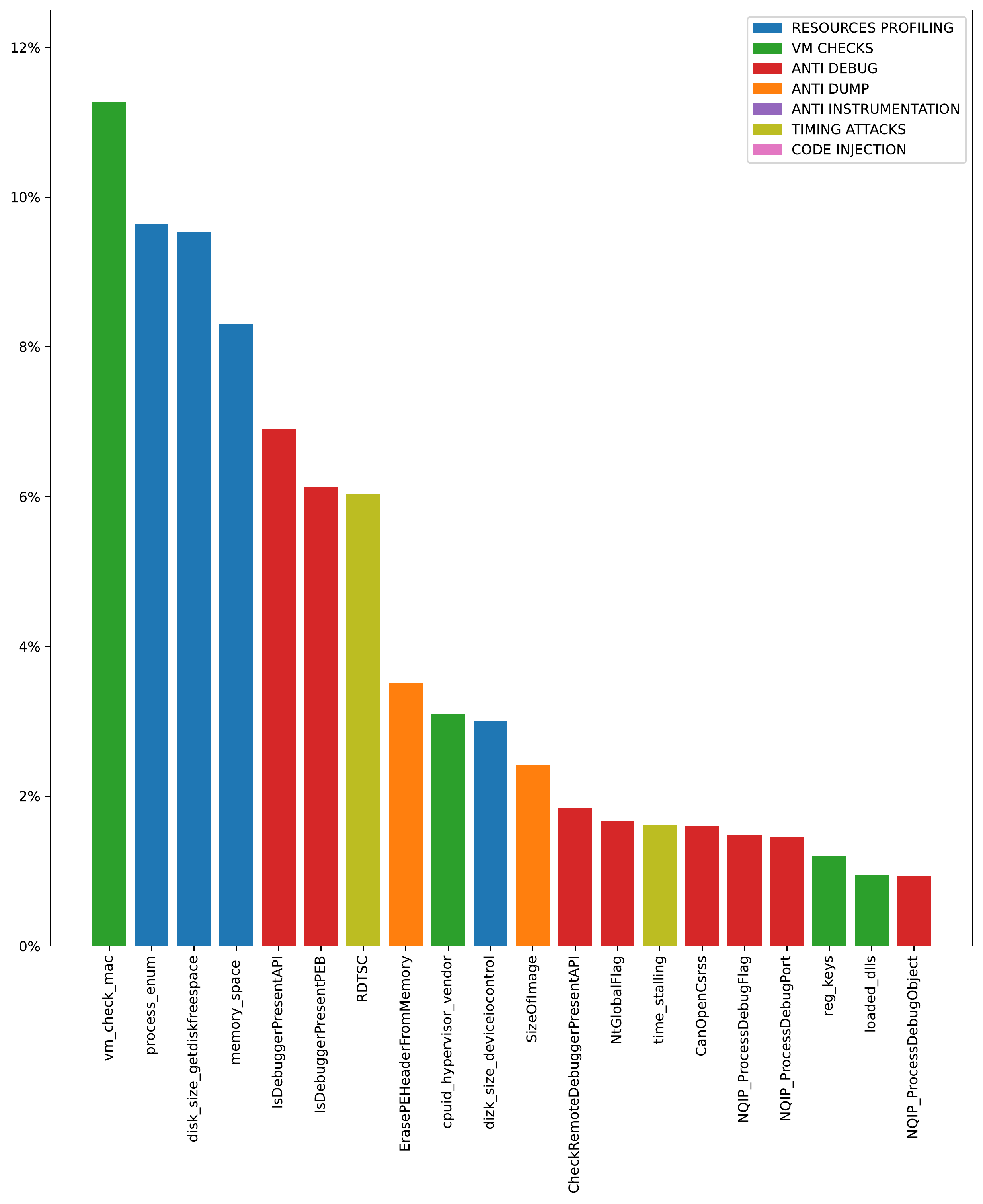}}\hfil
        \subfloat[2018]{\includegraphics[width=0.6\columnwidth]{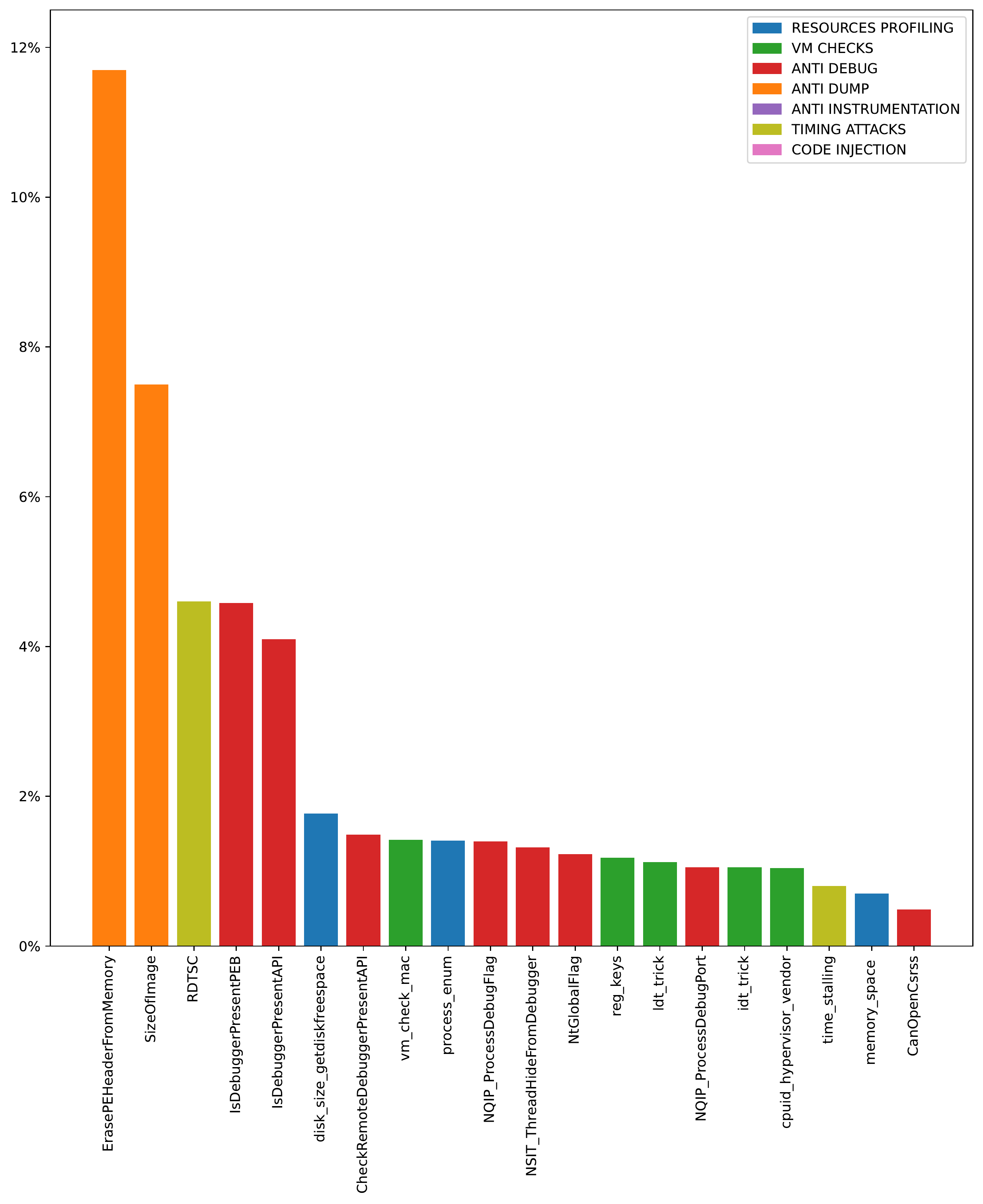}}\hfil
    
        \subfloat[2019]{\includegraphics[width=0.6\columnwidth]{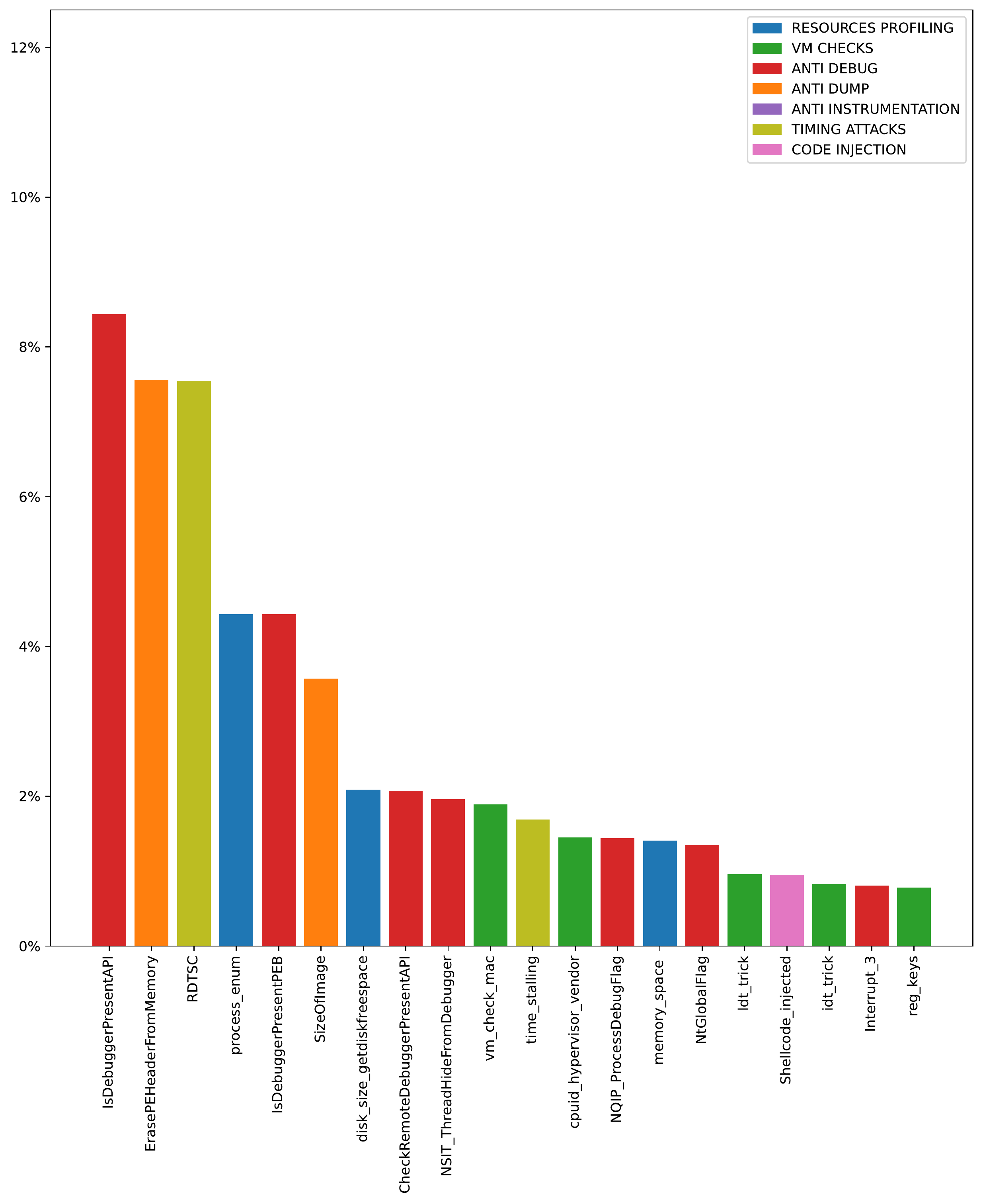}}\hfil
        \subfloat[2020]{\includegraphics[width=0.6\columnwidth]{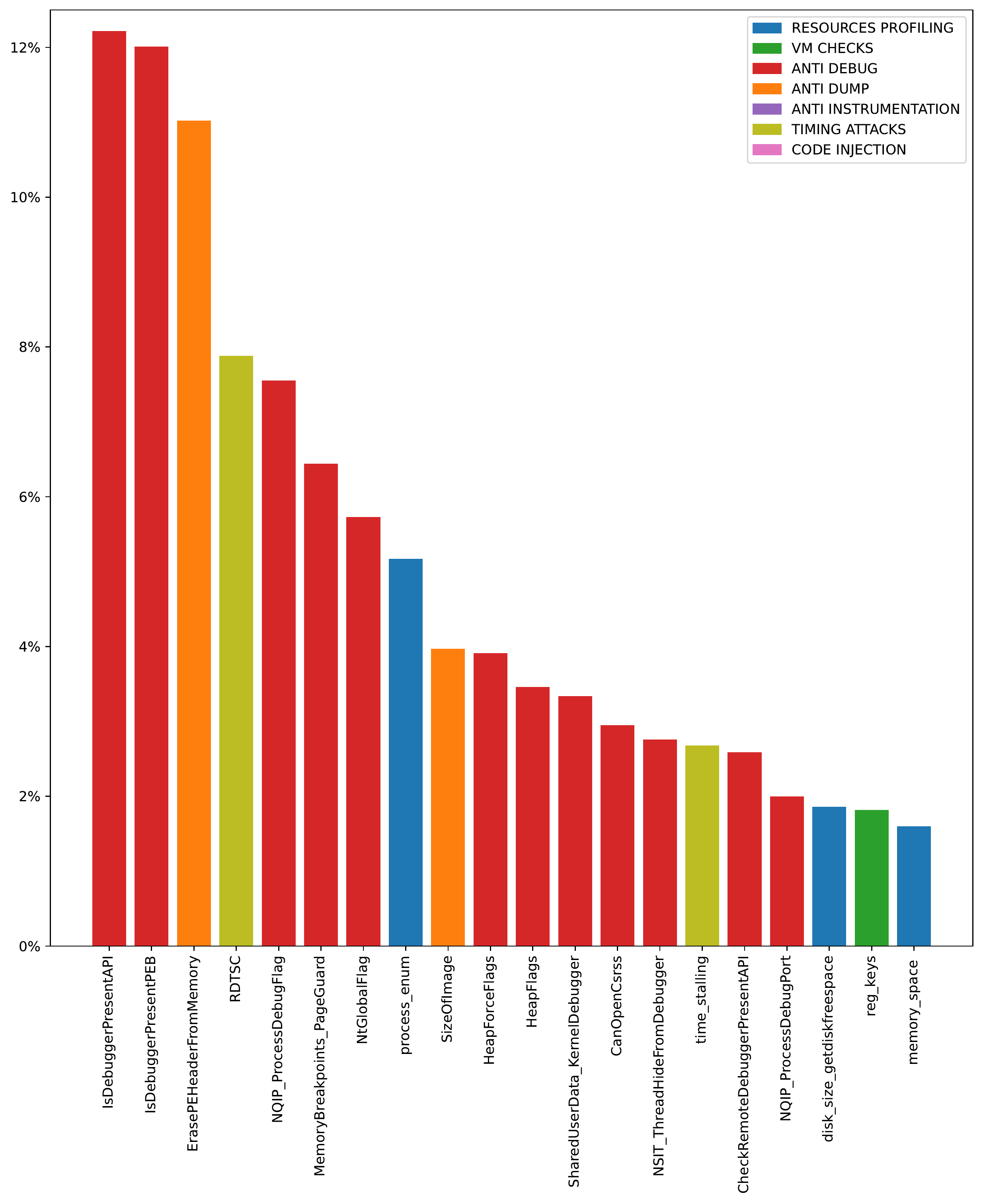}}\hfil
        \subfloat[Families]{\includegraphics[width=0.6\columnwidth]{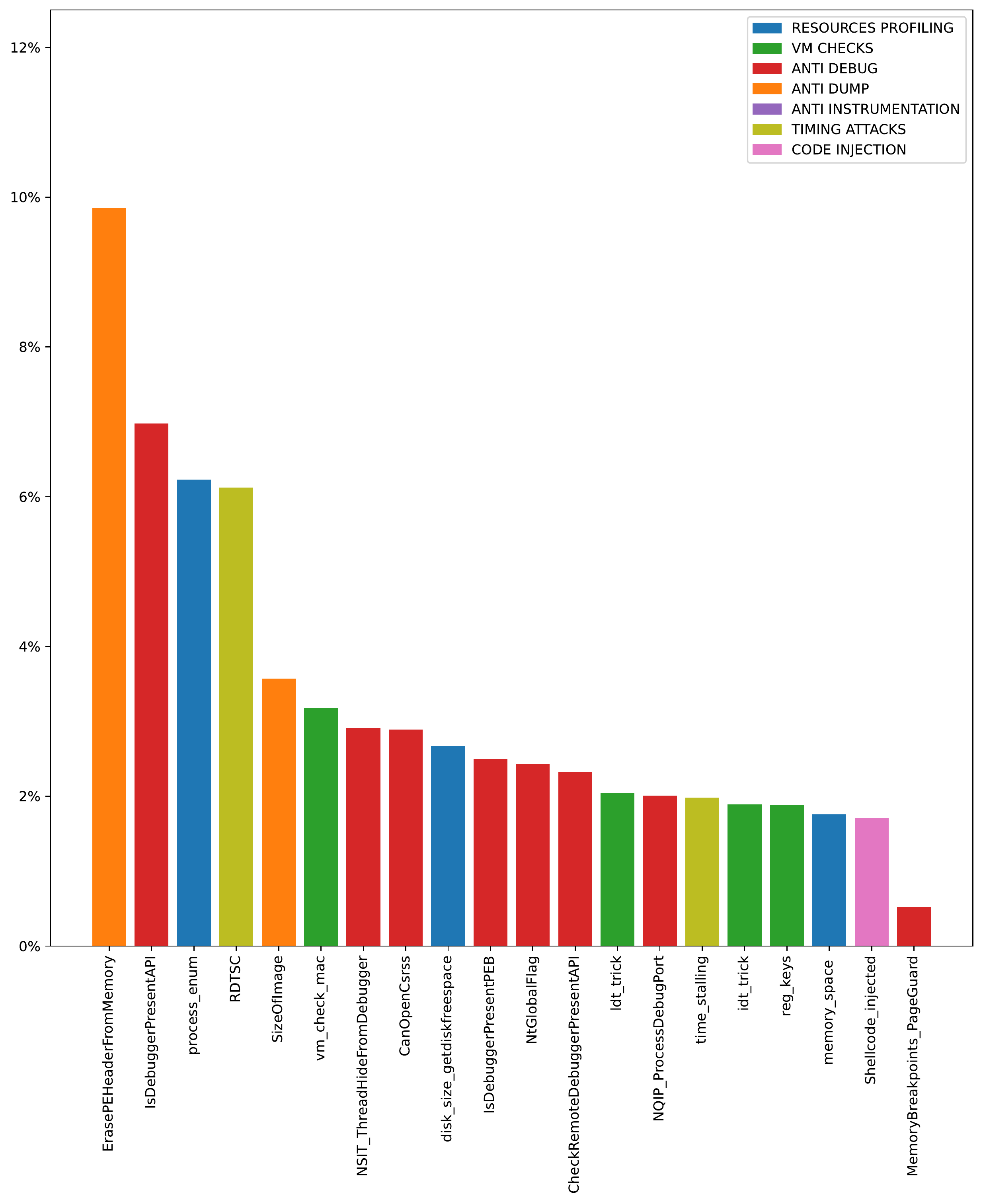}}
    \caption{Techniques prevalence over the years}
    \label{fig:techniques_prevalence_oty}

\end{figure*}

\begin{table*}[t]
    \centering
    \caption{Anti Dump Techniques}
    \label{tab:techniques_AntiDump}
    \scriptsize
    \begin{tabular}{m{4.5cm} m{6cm} m{3cm} m{3cm}}
    \toprule
    \textbf{Technique Name} & \textbf{Description} & \textbf{Detection} \\
    \bottomrule
    \texttt{ErasePEHeader} & Erase the PE Header signature inside process memory & Memory Write monitoring & \\ 
    \midrule
    \texttt{SizeOfImage} & Erase the SizeOfImage field inside the PE Header & Memory Write monitoring & \\
    \bottomrule
\end{tabular}
\end{table*}

\begin{table*}[t]
    \centering
    \caption{Code Injection Techniques}
    \label{tab:techniques_CodeInjection}
    \scriptsize
    \begin{tabular}{m{4.5cm} m{6cm} m{3cm} m{3cm}}
    \toprule
    \textbf{Technique Name} & \textbf{Description} & \textbf{Detection} & \textbf{Mitigation} \\
    \bottomrule
    \texttt{Shellcode\_injected} & Inject code inside another process using those APIs: \begin{itemize}
        \item NtWriteVirtualMemory
        \item NtCreateThreadEx
        \item NtResumeThread
        \item NtQueueApcThread
    \end{itemize} & API monitoring & Honeypot process \\
    \bottomrule
\end{tabular}
\end{table*}
\begin{table*}[t]
    \centering
    \caption{Timing Attacks Techniques}
    \label{tab:techniques_TimingAttack}
    \scriptsize
    \begin{tabular}{m{4.5cm} m{6cm} m{3cm} m{3cm}}
    \toprule
    \textbf{Technique Name} & \textbf{Description} & \textbf{Detection} & \textbf{Mitigation} \\
    \bottomrule
    \texttt{time\_stalling} & Wait a certain amount of time using \begin{itemize} \item {WaitForSingleObject} \item {TimeSetEvent} \item{NtDelayExecution} \item{SetWaitableTimer} \end{itemize} & API monitoring & Virtual Clock \\
    \midrule
    \texttt{RDTSC} & Check if time has been accellerated & Instruction monitoring & Sum to the returned value, our virtual clock \\
    \bottomrule
\end{tabular}
\end{table*}
% category, technique name, description, detection, mitigation
% detection = monitoring an instruction, an API, and or a memory access

\begin{table*}[ht]
\centering
\caption{Anti Debug Techniques}
\label{tab:techniques_AntiDebug}
\scriptsize
\begin{tabular}{ m{4.5cm} m{6cm} m{3cm} m{3cm} }
\toprule
\textbf{Technique Name} & \textbf{Description} & \textbf{Detection} & \textbf{Mitigation} \\
\bottomrule
IsDebuggerPresentAPI & Checks whether a debugger is present using \texttt{IsDebuggerPresent} & API monitoring & \\
\midrule
\texttt{IsDebuggerPresentPEB} & Checks whether a debugger is present checking \texttt{PEB->BeingDebugged} & Memory read monitoring & \\
\midrule
\texttt{CheckRemoteDebuggerPresentAPI} & Checks whether a debugger is present using \texttt{CheckRemoteDebuggerPresent} & API monitoring & \\
\midrule
\texttt{NSIT\_ThreadHideFromDebugger} & Hide a thread from the debugger using \texttt{NtSetInformationThread} with \texttt{ThreadHideFromDebugger} as parameter & API monitoring & \\
\midrule
\texttt{NtGlobalFlag} & Checks whether a debugger is present checking \texttt{PEB->NtGlobalFlag} & Memory read monitoring & \\
\midrule
\texttt{NQIP\_ProcessDebugPort} & Checks whether a debugger is present using \texttt{NtQueryInformationProcess} with \texttt{ProcessDebugPort} as ProcessInformationClass parameter& API monitoring & \\
\midrule
\texttt{NQIP\_ProcessDebugObject} & Checks whether a debugger is present using \texttt{NtQueryInformationProcess} with \texttt{ProcessDebugObject} as ProcessInformationClass parameter & API monitoring & \\
\midrule
\texttt{NQIP\_ProcessDebugFlag} & Checks whether a debugger is present using \texttt{NtQueryInformationProcess} with \texttt{ProcessDebugFlag} as ProcessInformationClass parameter & API monitoring & \\
\midrule
\texttt{CanOpenCsrss} & Checks whether the \texttt{SeDebugPrivileges} flag is active, trying to open Csrss.exe process & API monitoring & \\
\midrule
\texttt{MemoryBreakpoints\_PageGuard} & Checks whether a debugger is present throwing \texttt{STATUS GUARD PAGE VIOLATION}  & Memory access & If a guard page is accessed, trigger manually the exception \\
\midrule
\texttt{Interrupt\_0x2d} & Cause exception through \texttt{int 2d} instruction caught by the debugger if present. & Instruction monitoring & \\
\midrule
\texttt{Interrupt\_3} & Checks if a debugger catch the Breakpoint exception caused by \texttt{int 3} instruction & Instruction monitoring & \\
\midrule
\texttt{HardwareBreakpoints} & Checks whether any hardware breakpoint has been set & API monitoring & \\
\midrule
\texttt{NQSI\_SystemKernelDebuggerInformation} & Checks whether a debugger is present using \texttt{NtQuerySystemInformation} with \texttt{SystemKernelDebuggerInformation} as parameter & API monitoring & \\
\midrule
\texttt{HeapFlags} & Checks whether a debugger is present checking \texttt{PEB->ProcessHeap->Flags} & Memory read monitoring & \\
\midrule
\texttt{HeapForceFlags} & Checks whether a debugger is present checking \texttt{PEB->ProcessHeap->ForceFlags} & Memory read monitoring & \\
\midrule
\texttt{SharedUserData\_KernelDebugger} &  Checks whether a debugger is present reading \texttt{UserSharedData->KernelDebugger} & Memory read monitoring & \\
\midrule
\texttt{VirtualAlloc\_WriteWatch\_*} & Checks whether a debugger is present using \texttt{MEM\_WRITE\_WATCH} feature of \texttt{VirtualAlloc} & API monitoring & \\
\midrule
\texttt{NQO\_ObjectTypeInformation} & Checks whether a debugger is present calling \texttt{NtQueryObject} with parameter \texttt{ObjectTypeInformation}  & API monitoring & \\
\midrule
\texttt{NQO\_ObjectAllTypesInformation} & Checks whether a debugger is present calling \texttt{NtQueryObject} with parameter \texttt{ObjectAllTypesInformation} & API monitoring & \\
\midrule
\texttt{GetTickCount} & Checks whether a debugger is slowing down execution of the program & API monitoring & \\
\bottomrule
\end{tabular}
\end{table*}

\begin{table*}[t]
    \centering
    \caption{VM Checks Techniques}
    \label{tab:techniques_VMCheck}
    \scriptsize
    \begin{tabular}{m{4.5cm} m{6cm} m{3cm} m{3cm}}
    \toprule
    \textbf{Technique Name} & \textbf{Description} & \textbf{Detection} & \textbf{Mitigation} \\
    \bottomrule
    \texttt{reg\_keys} & Search hypervisor artifacts in registry keys & API monitoring &  \\
    \midrule
    \texttt{reg\_key\_value} & Search hypervisor artifacts in registry key values & API monitoring & \\
    \midrule
    \texttt{ldt\_trick} & Check Local Descriptor Table location with \texttt{sldt} instruction & Instruction monitoring & \\
    \midrule
    \texttt{idt\_trick} & Check Interrupt Descriptor Table location with \texttt{sidt} instruction & Instruction monitoring & \\
    \midrule
    \texttt{gdt\_trick} & Check Global Descriptor Table location with \texttt{sgdt} instruction & Instruction monitoring & \\
    \midrule
    \texttt{str\_trick} & Check Store Task Register with \texttt{str} instruction & Instruction monitoring & \\
    \midrule
    \texttt{vm\_check\_mac} & Check MAC address & API monitoring & \\
    \midrule
    \texttt{Firmware RSMB} & Get Firmware RSMB table and check for particular hypervisor artifact & API monitoring & Overwrite content of returned value \\
    \midrule
    \texttt{Firmware ACPI} & Get Firmware ACPI table and check for particular hypervisor artifact & API monitoring & Overwrite content of returned value \\
    \midrule
    \texttt{Device Artifacts} & Check for device artifacts created by various hypervisors & API monitoring &  \\
    \midrule
    \texttt{cpuid\_hypervisor\_vendor} & Get hypervisor vendor using \texttt{cpuid} with EAX=0x40000000 & Instruction monitoring & Modify the returned value inside EBX, ECX, EDX registers \\
    \midrule
    \texttt{cpuid\_is\_hypervisor} & Check the 31\textsuperscript{th} bit returned in ECX register by \texttt{cpuid} with EAX=0x1 & Instruction monitoring & Set 31\textsuperscript{th} bit to 0 \\
    \midrule
    \texttt{mouse\_movement} & Checks whether mouse position is always the same, calling multiple times \texttt{GetCursorPos} & API monitoring & Modify the returned value, with a randomly choosen one.\\
    \midrule
    \texttt{filesystem\_artifacts} & Check for particular hypervisor artifacts inside the filesystem & API monitoring & \\
    \midrule
    \texttt{setupdi\_diskdrive} & Check for hypervisor artifacts using \texttt{SetupDiGetDeviceRegistryPropertyW} & API monitoring & Zero out the returned buffer \\
    \midrule
    \texttt{manufacturer\_computer\_system\_wmi} & Get manufacturer through WMI & API monitoring & \\
    \midrule
    \texttt{model\_computer\_system\_wmi} & Get manufacturer through WMI & API monitoring & \\
    \midrule
    \texttt{vbox\_mac\_wmi} & Get vbox MAC address though WMI & API monitoring & \\
    \midrule
    \texttt{process\_id\_processor\_wmi} & Check ProcessId from Win32\_Processor using WMI & API monitoring & \\
    \midrule
    \texttt{serial\_number\_bios\_wmi} & Check SerialNumber devices using WMI & API monitoring & \\
    \bottomrule
\end{tabular}
\end{table*}
\begin{table*}[t]
    \centering
    \caption{Resource Profiling Techniques}
    \label{tab:techniques_Environment}
    \scriptsize
    \begin{tabular}{m{4.5cm} m{6cm} m{3cm} m{3cm}}
    \toprule
    \textbf{Technique Name} & \textbf{Description} & \textbf{Detection} & \textbf{Mitigation} \\
    \bottomrule
    \texttt{process\_enum} & Enumerate running processes through \texttt{Process32First/Process32Next} & API monitoring & Change parent process of the current process from pin.exe to cmd.exe\\
    \midrule
    \texttt{memory\_space} & Get RAM size using \texttt{GlobalMemoryStatusEx} & API monitoring & Return 8 GB of RAM \\
    \midrule
    \texttt{disk\_size\_getdiskfreespace} & Get hard disk size through \texttt{getdiskfreespaceEx} & API monitoring & Return 800GB of Hard disk \\
    \midrule
    \texttt{dizk\_size\_deviceiocontrol} & Checks hard disk size using \texttt{DeviceIoControl} & API monitoring & Return 800GB of Hard disk \\
    \midrule
    \texttt{disk\_size\_wmi} & description & API monitoring & Disabled WMI service \\
    \midrule
    \texttt{NumberOfProcessors} & Checks whether the current machine has less than 2 processors & Read memory monitoring & Return 4 processors \\
    \bottomrule
\end{tabular}
\end{table*}

\begin{table*}[t]
    \centering
    \caption{Anti Instrumentation Techniques}
    \label{tab:techniques_AntiInstrumentation}
    \scriptsize
    \begin{tabular}{m{4.5cm} m{6cm} m{3cm} m{3cm}}
    \toprule
    \textbf{Technique Name} & \textbf{Description} & \textbf{Detection} & \textbf{Mitigation} \\
    \bottomrule
    \texttt{Check\_EIP} & Leak Instruction pointer using: 
    \begin{itemize}
        \item FPU instructions
        \item Int 0x2e
    \end{itemize} & Instruction monitoring & Return the Instruction pointer expected by the process, hiding code cache \\
    \bottomrule
\end{tabular}
\end{table*}

\end{document}